\documentclass[
  journal=largetwo,
  manuscript=article-type,
  year=2020,
  volume=37,
]{cup-journal}
\usepackage{tikz}
\usepackage{amssymb}
\usepackage{amsmath}
\usepackage{mathtools}
\usepackage[utf8]{inputenc}  
\usepackage[T1]{fontenc}  
\usepackage[nopatch]{microtype}
\usepackage{booktabs}
\usepackage[colorlinks=true,linkcolor=purple, citecolor=purple, allcolors=blue]{hyperref}
\usepackage{xcolor}
\DeclareUnicodeCharacter{2248}{$\approx$}

\usetikzlibrary{shapes.geometric, arrows}
\tikzstyle{process} = [rectangle, minimum width=3em, minimum height=2em, text centered, draw=blue, fill=gray!10]
\tikzstyle{startend} = [ellipse, minimum width=2em, minimum height=1em, text centered, draw=red, fill=gray!10]
\tikzstyle{arrow} = [thick,->,>=stealth]

\title{21cm Epoch of Reionisation Power Spectrum with Closure Phase using the Murchison Widefield Array}

\author{Himanshu Tiwari}
\affiliation{International Centre for Radio Astronomy Research (ICRAR), Curtin University, Kent Street, Bentley, Perth, Western Australia, 6102.}
\alsoaffiliation{Commonwealth Scientific and Industrial Research Organisation (CSIRO), Space \& Astronomy, P. O. Box 1130, Bentley, WA 6102, Australia.}
\email{himanshu.tiwari@postgrad.curtin.edu.au}

\author{Nithyanandan Thyagarajan}
\affiliation{Commonwealth Scientific and Industrial Research Organisation (CSIRO), Space \& Astronomy, P. O. Box 1130, Bentley, WA 6102, Australia.}

\author{Cathryn M. Trott}
\affiliation{International Centre for Radio Astronomy Research (ICRAR), Curtin University, Kent Street, Bentley, Perth, Western Australia, 6102.}
\alsoaffiliation{ARC Centre of Excellence for All Sky Astrophysics in Three Dimensions (ASTRO-3D), Australia.}

\author{Benjamin McKinley}
\affiliation{International Centre for Radio Astronomy Research (ICRAR), Curtin University, Kent Street, Bentley, Perth, Western Australia, 6102.}
\alsoaffiliation{ARC Centre of Excellence for All Sky Astrophysics in Three Dimensions (ASTRO-3D), Australia.}


\begin{document}

\begin{abstract}
The radio interferometric closure phases can be a valuable tool for studying cosmological {H\scriptsize{I}}~from the early Universe. Closure phases have the advantage of being immune to element-based gains and associated calibration errors. Thus, calibration and errors therein, which are often sources of systematics limiting standard visibility-based approaches, can be avoided altogether in closure phase analysis. In this work, we present the first results of the closure phase power spectrum of {H\scriptsize{I}}~21-cm fluctuations using the Murchison Widefield Array (MWA), with $\sim 12$ hours of MWA-phase II observations centered around redshift, $z\approx 6.79$, during the Epoch of Reionisation. On analysing three redundant classes of baselines -- 14~m, 24~m, and 28~m equilateral triads, our estimates of the $2\sigma$ ($95\%$ confidence interval) 21-cm power spectra are  $\lesssim (184)^2 pseudo \rm ~mK^2$ at ${k}_{||} = 0.36 $ $pseudo~h {\rm Mpc^{-1}}$ in the EoR1 field for the 14~m baseline triads, and $\lesssim (188)^2 pseudo \rm ~mK^2$ at $k_{||} = 0.18 $ $pseudo~h {\rm Mpc^{-1}}$ in the EoR0 field for the 24~m baseline triads. The ``$pseudo$'' units denote that the length scale and brightness temperature should be interpreted as close approximations. Our best estimates are still 3-4 orders high compared to the fiducial 21-cm power spectrum; however, our approach provides promising estimates of the power spectra even with a small amount of data. These data-limited estimates can be further improved if more datasets are included into the analysis. The evidence for excess noise has a possible origin in baseline-dependent systematics in the MWA data that will require careful baseline-based strategies to mitigate, even in standard visibility-based approaches.
 \end{abstract}

\section{Introduction} \label{section: Introduction}

Epoch of Reionisation (EoR) is the period when the first stars and galaxies were formed in the early Universe between $15<z<5.3$ and contributed to reionising the predominantly neutral intergalactic medium on cosmic scales \citep{Bosman_2022, Zhu_2022, Zhu_24}. It is also one of the least understood periods in the history of the Universe, mainly due to the lack of radiation influx from the first stars and galaxies, which are locally absorbed by the intervening medium. The hyperfine ground state of the atomic Hydrogen (H{\scriptsize{I}}) produces a weak transition of $\sim 1420$ MHz, popularly known as the 21-cm line. It is considered a very promising probe of the EoR due to the abundance of Hydrogen in the early Universe. The intervening medium is largely transparent to the redshifted 21-cm line; therefore, it provides one of the best avenues to infer the astrophysical properties of the IGM and the cosmology of the early Universe. As the neutral IGM gets ionised, it weakens the strength of the 21-cm signal. 
One can interpret the stages of cosmic reionisation by estimating the depletion in the redshifted 21-cm signal through cosmic time \citep[see][for review]{Furlanetto_2004, Pritchard_2012, Mesinger_2016}.\par
To detect this forbidden transition from the early Universe, several radio instruments such as Murchison Widefield Array (MWA) \citep{Tingay_13},  Hydrogen Epoch of Reionization Array (HERA) \citep{deboer17}, LOw-Frequency ARray (LoFAR) \citep{Haarlem_2013}, Giant metrewave Radio Telescope (GMRT) \citep{Paciga13}, Precision Array for Probing the Epoch of Reionization (PAPER) \citep{Pober_2011}, Long Wavelength Array (LWA) \citep{Eastwood_2019}, Experiment to Detect the Global EoR Signature (EDGES) \citep{Bowman_2008, Bowman_2010, Bowman_18}, Shaped Antenna measurement of the background RAdio Spectrum (SARAS) \citep{patra_2012,singh_2018, saras3_singh2022}, Broad-band Instrument for Global HydrOgen ReioNization  Signal (BIGHORNS) \citep{Sokolowski_2015}, Large Aperture Experiment to Detect the Dark Age (LEDA) \citep{Bernardi_2016}, Dark Ages Radio Explorer (DARE) \citep{Burns_2012},  Sonda Cosmol\'ogica de las Islas para la Detecci\'on de
Hidr\'ogeno Neutro (SCI-HI) \citep{Voytek_2014}, Probing Radio Intensity at High-Z from Marion (PRIZM) \citep{prizm_17} were built or are under construction. These instruments can either aim to detect the sky-averaged 21-cm signal spectrum (Global signal) or measure its spatial fluctuations. The former category of instruments can detect the overall IGM properties, whereas the latter can provide a detailed study of the three-dimensional topology of the EoR regime. 
The spatial signatures are quantified through statistical measures such as the power spectrum, which can probe the 21-cm signal strength as a function of cosmological length scales ($k$-modes). Alongside, the three-dimensional topology of the EoR can be studied via a two-dimensional 21-cm power spectrum \citep{barry19, Trott_2020, Mertens_2020, HERA_2021, munshi2023upper}, which shows the variation of the 21-cm power spectrum along the line of sight and transverse axis.\par
The significant challenges of detecting the 21-cm signal come from the foregrounds, ionospheric abnormalities, instrumental systematics, and Radio Frequency Interference (RFI), emitting in the same frequency range as the redshifted 21-cm signal from the EoR. In an ideal situation avoiding or minimising all the above factors, we still require to calibrate the instrument against the bright foregrounds that require calibration accuracy of $\gtrsim 10^5 : 1$ by the radio instruments to reach the required H{\scriptsize{I}}~levels. 
Calibration accuracy is especially important for EoR observations using the MWA because of the presence of sharp periodic features in the bandpass produced by the polyphase filter bank used. The inability to accurately correct for these element-based bandpass structures significantly affects the power spectrum estimates \citep{Beardsley+2016,barry19, Trott_2020, Patwa+2021, Yoshiura_2021}. \par
The radio-interferometric closure phase has emerged as an alternative and independent approach to studying the EoR while addressing the calibration challenges. The main advantage of this approach is the immunity of closure phases to the errors associated with the direction-independent, antenna-based gains. Thus, calibration is not essential in this approach \citep{Carilli_18, Nithya_20a}. The closure phase in the context of the EoR was first investigated by \citet{Nithya_18, Carilli_18} and further employed on the HERA data by \citet{Nithya_20b, Keller_2023}. The method has shown significant promise in avoiding serious calibration challenges, and with detailed forward modelling, one can ideally quantify the 21-cm power spectrum. This paper is the first attempt to utilise a closure phase approach on MWA-phase II observations. We followed the methods investigated by \citet{Nithya_20b, Keller_2023} and applied them to our datasets. This paper is organised as follows. In \S{\ref{section: background}}, we discuss the background of the closure phase. \S{\ref{section: observation}} and \S{\ref{section: simulations}} of this paper explain the observations and forward modeling with simulations of the foregrounds, H{\scriptsize{I}}, and noise. In \S{\ref{section: data_processing}}, we discuss the data processing and rectification, and finally, we present our results in \S{\ref{section: results}} and discuss them in \S{\ref{section: discussion}}.

\section{Background} \label{section: background}

In this section, we review the background of the interferometric closure phase in brief \citep[refer to][for a complete mathematical understanding of this approach]{Nithya_18, Nithya_20a}. The measured visibility between two antenna factors at a given baseline $(V_{ij}^{\rm m})$ can be defined as the sum of true sky visibility and noise:
\begin{equation}\label{eq:v_ij}
    V_{ij}^{\rm m}(\nu) = \mathbf{g}_{i}(\nu) V_{ij}^{\rm T}(\nu)\mathbf{g}_{j}^*(\nu)  + V_{ij}^{\rm N}(\nu);
\end{equation}
where $\mathbf{g}_{i}(\nu), ~\mathbf{g}_{j}(\nu)$ denote the element-based gain terms, \{$*$\} represents the complex conjugate, $V_{ij}^{\rm T}(\nu)$ is the true sky visibility, and $V_{ij}^{\rm N}(\nu)$ is the noise in the measurement.  The indices $\{ij\}$ correspond to the antennae $\{a,b\}$ forming a baseline. The true sky visibility can be further decomposed into the foregrounds and faint cosmological H{\scriptsize{I}}~visibilities. 
\begin{equation}
    V_{ij}^{\rm T}(\nu) = V_{ij}^{\rm FG}(\nu) + V_{ij}^{\rm H\scriptsize{I}~}(\nu)
\end{equation}
In general, the foreground $\geq 10^4\rm K$ orders of higher magnitude than the H{\scriptsize{I}}, and to reach the sensitivity limit of H{\scriptsize{I}}, the gains $(\mathbf{g}_i's)$ are required to be precisely calibrated up to the H{\scriptsize{I}}~levels. It presents challenges to the direct visibility-based H{\scriptsize{I}}~power spectrum analysis as it requires accurate modelling of the foreground and mastering the calibration techniques.
In radio interferometry, the term `closure phase' is assigned to the phase derived from the product of $N \geq 3$ closed loops of antenna visibilities \citep{jennison58}. When $N=3$, it is also sometimes referred to as the bispectrum phase in the literature, which can be defined as,

\begin{align}\label{eq: closure_phase}
    \phi_\nabla^{\rm m}(\nu) &= {\rm arg} \prod_{ij=1}^3 V_{ij}^{\rm m}(\nu)  \nonumber \\
   &=  {\rm arg} \prod_{ij=1}^3 \left[ \mathbf{g}_{i}(\nu) V_{ij}^{\rm T}(\nu)\mathbf{g}_{j}^*(\nu)  + V_{ij}^{\rm N}(\nu) \right]  \nonumber \\
   &= {\rm arg} \prod_{ij=1}^3 V_{ij}^{\rm T}(\nu) + \textrm{noise-like terms} \,
\end{align}
where, $\{ij\}$ runs through antenna-pairs $\{ab, bc, ca\}$, the gains of individual antenna elements $(\mathbf{g}_{i}'\rm s)$ gets eliminated in the closure phase; leaving only the true sky phase as the sole contributing factor in the closure phase. \par
The closure phase delay spectrum technique also exploits the fact that the foregrounds predominantly obey a smooth spectral behavior, whereas the cosmological H{\scriptsize{I}}~creates spectral fluctuations. Thus, in the Fourier delay domain, the foreground signal strength (power) gets restricted within the lower delay modes. In contrast, the H{\scriptsize{I}}~power can be observed at the higher delay modes, creating the distinction between these two components in the Fourier domain, from which the faint H{\scriptsize{I}}~can be detected. Because of its advantages in avoiding element-based calibration errors and processing simplicity, it promises to be an independent alternative to estimating the 21-cm power spectrum \citep{Nithya_18, Carilli_18, Nithya_20a, Nithya_20b, Keller_2023}. \par
The delay spectrum of the closure phase can be estimated by taking the Fourier transform of the complex exponent of the closure phase with a window function along frequency,
\begin{equation}\label{eq:psi_tau}
    \tilde\Psi_\nabla(\tau) = V_{\rm eff} \int e^{i\phi_\nabla(\nu)} W(\nu) e^{2\pi i \nu \tau} d\nu \, ,
\end{equation}
where $\tau$ represents delay, the Fourier dual of the sampling frequencies, and $V_{\rm eff}$ is the effective visibility, which can be obtained through the model foreground visibilities or a calibrated visibility. In our work, we used the former estimated through foreground simulations, which are discussed in the next section. Note that we only take the amplitude of the $V_{\rm eff}$, which acts as a scaling factor in the delay spectrum. $W(\nu)$ is the spectral window function; we used a Blackman-Harris window \citep{Harris_1978} modified to obtain a dynamic range required to sufficiently suppress foreground contamination in the EoR window \citep{Nithya_2016},
\[W(\nu) = W_{\rm BH}(\nu) \ast W_{\rm BH}(\nu) \, ,\] 
where, $W_{\rm BH}(\nu)$ is the Blackman-Harris window function, and $\{\ast\}$ represents the convolution operation. For a given triad, $V_{\rm eff}$ is estimated from the sum of inverse variance visibilities weighted over the window,
\begin{equation}\label{eq: veff}
    V_{\rm eff}^{-2} = \sum_{ij=1}^3 \frac{\int V_{ij}(\nu)^{-2} W(\nu) d\nu}{\int W(\nu) d\nu} \, .
\end{equation}
The inverse squaring ensures the appropriate normalisation of visibilities, $V_{ij}$, taking noise into account. From the delay spectrum, we estimate the delay cross power spectrum by taking the product of the two delay spectra and converting it into [$pseudo~\textrm{mK}^2 h^{-3} \textrm{Mpc}^3$] units by assimilating the cosmological and antenna-related factors as,
\begin{multline}\label{eq: power_kpar}
     P_\nabla(k_{||}) = 2\mathcal{R}\{\tilde\Psi_{\nabla}(\tau)*{\overline{\tilde\Psi_{\nabla}^\prime(\tau)}}\} \times \\
            \left(\frac{1}{\Omega B_{\rm eff}}\right)\left(\frac{D^2\Delta D}{B_{\rm eff}}\right)
           \left(\frac{\lambda^2}{2k_B}\right)^2  
           [pseudo~ {\rm mK}^2h^{-3}{\rm Mpc^3}] \, ,   
\end{multline}
where $\Omega$ is the antenna beam squared solid angle \citep{Parsons_2014}, 
$B_{\rm eff}$ is the effective bandwidth of the observation, $D$ and $\Delta D$ are the cosmological comoving distance and comoving depth corresponding to the central frequency and the bandwidth, respectively. $k_{||}$ is the wavenumber along the line-of-sight \citep{Morales2004},
\begin{equation}
    k_{||} = \frac{2\pi\tau B_\textrm{eff}}{\Delta D} \approx \frac{2\pi\tau \nu_r H_0 E(z)}{c(1+z)^2} \, ,
\end{equation}
where, $\nu_r$ is the redshifted 21-cm frequency, $H_0$ and $E(z)$ are standard terms in cosmology. 

$\Omega B_{\rm eff}$ is related to the cosmological volume probed by the instrument and is defined as \citep{Nithya_2016}
\begin{align}
    \Omega B_{\rm eff} &= \iint |A(\hat{\mathbf{s}},\nu)|^2 \, |W(\nu)|^2 \, \mathrm{d}^2 \hat{\mathbf{s}} \, \mathrm{d}\nu \, ,
\end{align}
where $A(\hat{\mathbf{s}},\nu)$ is the frequency-dependent, directional power pattern of the antenna pair towards the direction, $\hat{\mathbf{s}}$, and $W(\nu)$ is the spectral window function. However, we approximated by separating the integrals to obtain $\Omega$ as \citep{Parsons_2014}
\begin{align}
    \Omega &= \int |A(\hat{\mathbf{s}},\nu_r)|^2 \, \mathrm{d}^2 \hat{\mathbf{s}}
\end{align}
and effective bandwidth, $B_\textrm{eff}$, as \citep{Nithya_2016}
\begin{align}
    B_\textrm{eff} &= \epsilon B = \int_{-B/2}^{+B/2} |W(\nu)|^2 \mathrm{d}\nu \, ,
\end{align}
where, $\epsilon$ is the spectral window function's efficiency, and $B=30.72$~MHz is MWA's instantaneous bandwidth. For the MWA observations at the chosen band in this study, $\Omega\approx 0.076$~Sr. For the modified Blackman-Harris window function adopted here, $\epsilon\approx 0.42$ and hence, $B_\textrm{eff} \approx 12.90$~MHz. 

The ``$pseudo$'' in Equation~(\ref{eq: power_kpar}) is used to note that the power spectrum estimated via the closure phase method is an approximate representation of the visibility-based power spectrum \citep{Nithya_20a}. Further, we used the power spectrum as defined in \cite{Nithya_20b} with the scaling factor $2$ instead of $2/3$\footnote{The definition of $V_{\rm eff}^2$ already accounts for the factor of 3} to correct for the effective visibility estimates.

\section{Observations} \label{section: observation}

\begin{figure}
    \centering
    \includegraphics[scale=0.45]{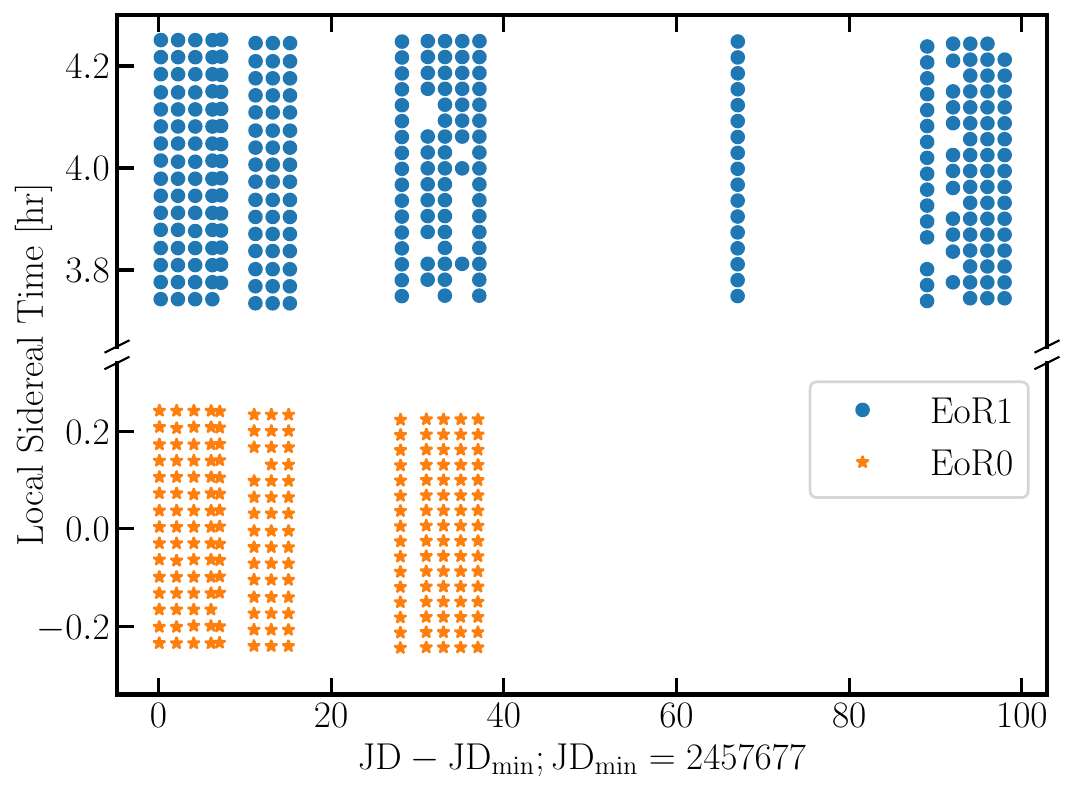}
    \caption{observations used in this analysis: \textit{blue} circles and \textit{orange} stars represent the individual observations made in the respective EoR fields.}
    \label{fig: obs_info}
\end{figure}

In this work, we used 493 zenith-pointed MWA-phase II compact observations from September 2016 under the MWA project \texttt{EoR-HighSeason}. Each observation lasts 112 seconds in the MWA high-band frequency range of 167-197 MHz. Amongst these observations, 198 and 295 target the EoR0 (RA, Dec = 0h, -27deg) and EoR1 (RA, Dec = 4h, -30deg) fields, respectively. Figure \ref{fig: obs_info} shows the Local Sidereal Time (LST) and Julian Date (JD) of the observations. It amount to $\approx$~6 and 9 hours of total observing time on the EoR0 and EoR1 fields, respectively. The MWA raw visibilities are stored in the \texttt{gpu-fits} format. To convert them into measurement sets (MS) or uvfits, we used \texttt{Birli}\footnote{\url{https://github.com/MWATelescope/Birli}} (an MWA-specific software that can perform data conversion, averaging in frequency and time, flagging, and other preprocessing steps). Using \texttt{Birli}, we averaged the raw visibilities for 8 seconds at a frequency resolution of 40 kHz. 
Finally, we output the raw (uncalibrated) visibilities as standard \texttt{uvfits}. \par
Note that we are required to keep all frequency channels for the closure phase analysis; thus, we avoided flagging channel-based RFI (e.g., DTV) and coarse band edge channels (around every 1.28 MHz), which are usually affected by the bandpass\footnote{The MWA’s signal processing chain contains filterbanks that yield 24 coarse channels of 1.28 MHz over the full 30.72 MHz band. The fine polyphase filterbank shape results in poor bandpass characteristics at the coarse channel edges \citep{Trott_2020}.}. 

\section{Simulations} \label{section: simulations}

The closure phases are not linear in the visibilities; thus, forward modelling is key to understanding the data. We incorporated simulations of the foregrounds (FG), H{\scriptsize{I}}, and antenna noise to provide cross-validation and comparison with the data. Forward modelling can help identify the excess noise and systematic biases in the data and provide an idealistic estimate for comparison.

\begin{figure}
    \centering
    \includegraphics[scale=0.45]{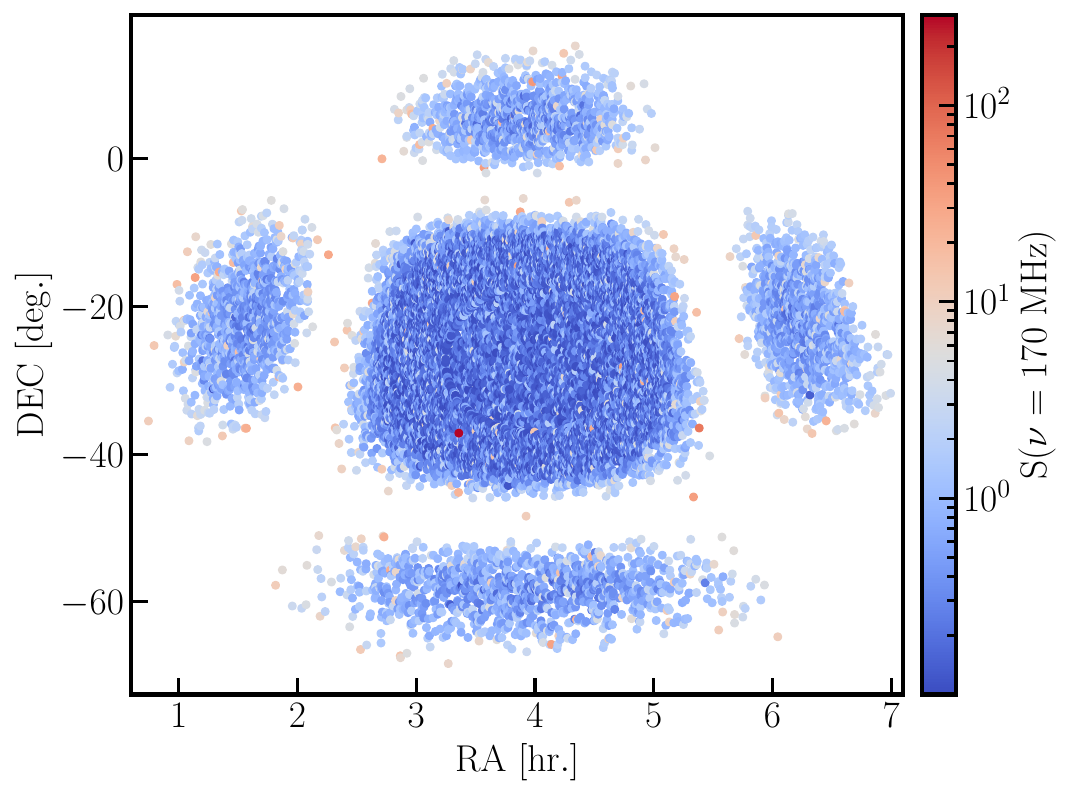}
    \caption{Beam attenuated sky-map of 20,000 sources in the EoR1 field used in the foreground simulation. The corresponding Stokes-I flux density at 170 MHz is shown with the color scale. The sources are only shown as point sources (single component) in the above figure. }
    \label{fig: foreground-sources}
\end{figure}

\subsection{foregrounds} \label{subsection: foreground_simulation}

The simulated FG are generated with the same parameters (i.e. matching LST, frequency and time resolution) as the observing data. In the first step, we generated the sky maps corresponding to each individual observation. We used the top 20,000 brightest radio sources from the PUMA catalogue \citep{line_2017} in the observing fields (EoR0, EoR1).
The catalogue includes point sources, Gaussians and shapelets. Note that our sky model does not account for the diffuse sky emission. Then, we generated the foreground sky visibilities and converted them to MWA-style \texttt{uvfits}. Initially, we experimented with various source counts (e.g., 15000, 25000, 45000) and their effect on the closure phase. We found the variation in closure phases saturates beyond 20000. Therefore, we settled for 20000 source counts in favour of faster computation. However, it should be noted that pinpointing the exact number of source counts where the closure phase saturates is challenging to find. The entire task of sky-map generation and foreground visibility estimation was accomplished using \texttt{Hyperdrive}\footnote{\url{https://github.com/MWATelescope/mwa_hyperdrive}}. The sky visibilities are generated using fully embedded antenna element (FEE) beam with \textit{real}-MWA observing scenario where the information of dead dipoles (if present during the observation) and antenna gains are incorporated in the simulation. Figure \ref{fig: foreground-sources} shows 20,000 sources around the EoR1 field for a given observation. For simplicity, only the single component of the sources (or point sources flux density)  is shown in the figure.

\subsection{Neutral hydrogen}\label{subsection: HI_simulation}

\begin{figure}
    \centering
    \includegraphics[scale=0.4]{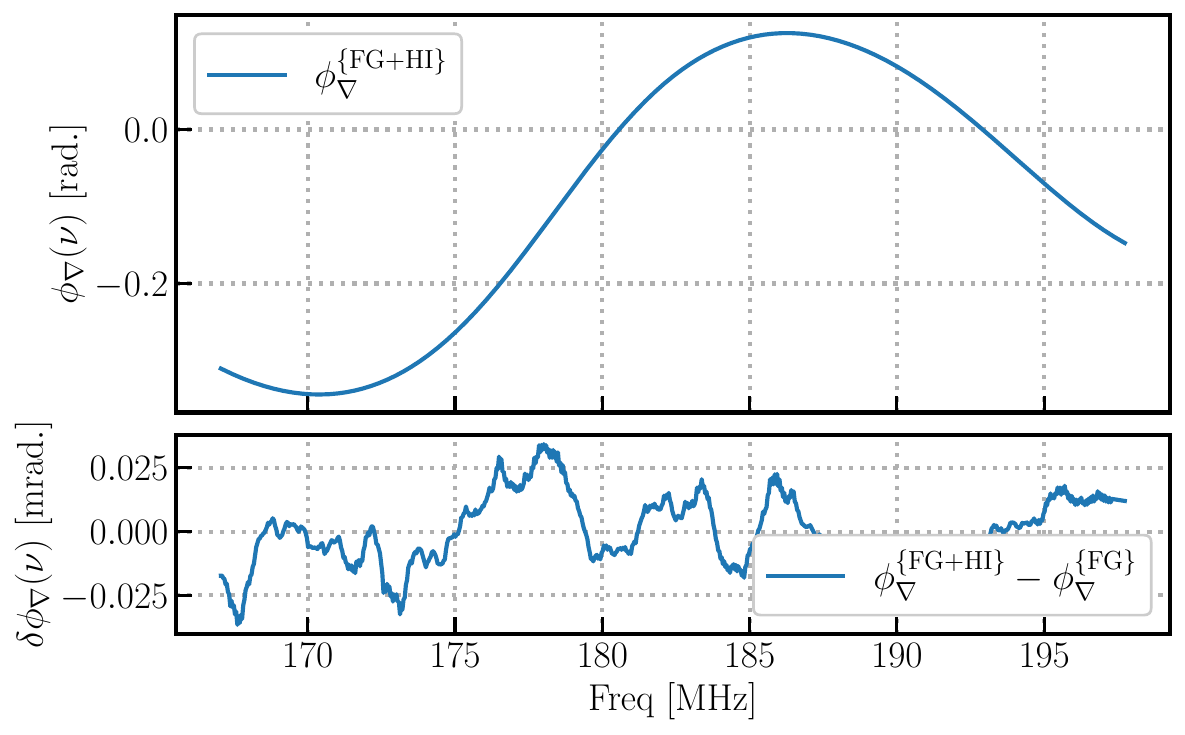}
    \caption{Top: closure phase of the sum of the visibilities of the foreground and H{\scriptsize{I}}~simulation {\{FG+{H\scriptsize{I}}\}} for a single triad of 14~m baseline length. Bottom: the difference between the closure phases of \{{FG+{H\scriptsize{I}}\}} and FG-only simulation, showing the sub-milliradian level fluctuation of the embedded H{\scriptsize{I}}-signal in the closure phase.}
    \label{fig: closure-phase}
\end{figure}

Next, we estimate the H{\scriptsize{I}}~visibilities as observed by MWA. In the limits of the cosmic and sample variance, the characteristic fluctuations in the H{\scriptsize{I}}-signal can be assumed to be the same across the sky; therefore, we can avoid simulating H{\scriptsize{I}}~box multiple times; instead, we can use a single H{\scriptsize{I}}~simulation box. The H{\scriptsize{I}}~simulation was generated using 21cmFAST \citep{Mesinger_2010} with a simulation box size of $1.5$~cGpc corresponding to $50^\circ \times  50^\circ$ in the sky at a redshift of 6. Then, we passed the simulated voxel data cube to \texttt{WODEN}\footnote{\url{https://github.com/JLBLine/WODEN}} \citep{Line2022} to generate the MWA-style visibilities of the H{\scriptsize{I}}~and output it as \texttt{uvfits}. The H{\scriptsize{I}}~visibilities were first generated for each $1.28$ MHz coarse band separately with the matching frequency and time resolution of $40$ kHz and 8 seconds of the foreground simulations and then manually stitched together to get the total 30.72 MHz bandwidth. The final H{\scriptsize{I}}~visibilities were passed to the processing pipeline for further analysis.
The foreground and H{\scriptsize{I}}~visibilities are added together. We computed the closure phase spectra of the foregrounds as well as of the H{\scriptsize{I}}~imprinted on the foregrounds. Figure~\ref{fig: closure-phase} shows the smooth foreground spectra in the closure phase and the fluctuations ($\sim 0.01$~milliradian) introduced by the presence of H{\scriptsize{I}}. 

\subsection{Noise} \label{subsection: Noise_simulation}

The total noise consists of sky noise and receiver noise components. 
The receiver temperature for the MWA was assumed to be $T_{\rm rx}=50\rm ~K$ \citep{Daniel2020}, while sky temperature follows a power law in our observing frequency range \citep{PReport},
\[T_{\rm sky} = T_0\left(\frac{\nu}{\nu_0}\right)^\alpha; T_0=180~{\rm K}, \nu_0=180~{\rm MHz}, \alpha=-2.5 \, ,\]
\[T_{\rm sys} = T_{\rm sky} + T_{\rm rx} \, .\]
From the system temperature, we estimated the system-equivalent flux density (SEFD) using \citet{Thompson_2017},
\[{\rm SEFD} = \frac{2k_B T_{\rm sys}}{A_{\rm eff}}\]
and the RMS,
\begin{equation}
     {\sigma (\nu)} = \frac{{\rm SEFD}}{\sqrt{\Delta\nu\Delta t}} \, ,
\end{equation}
where, $k_B$ is Boltzmann's constant, $A_{\rm eff}$ is the effective collecting area of the telescope, and $\Delta \nu, \Delta t$ are the frequency resolution and integration time, respectively.  The ${\sigma(\nu)}$ is used to generate the Gaussian random noise and converted into the complex noise visibilities with a normalisation factor of $1/\sqrt{2}$ in the real and imaginary parts. Finally, the noise visibilities were added to the corresponding foreground and H{\scriptsize{I}}~visibilities to get the Model of the sky signal.

\subsection{Baseline-dependent gains}
\begin{figure}
    \centering
    \includegraphics[scale=0.4]{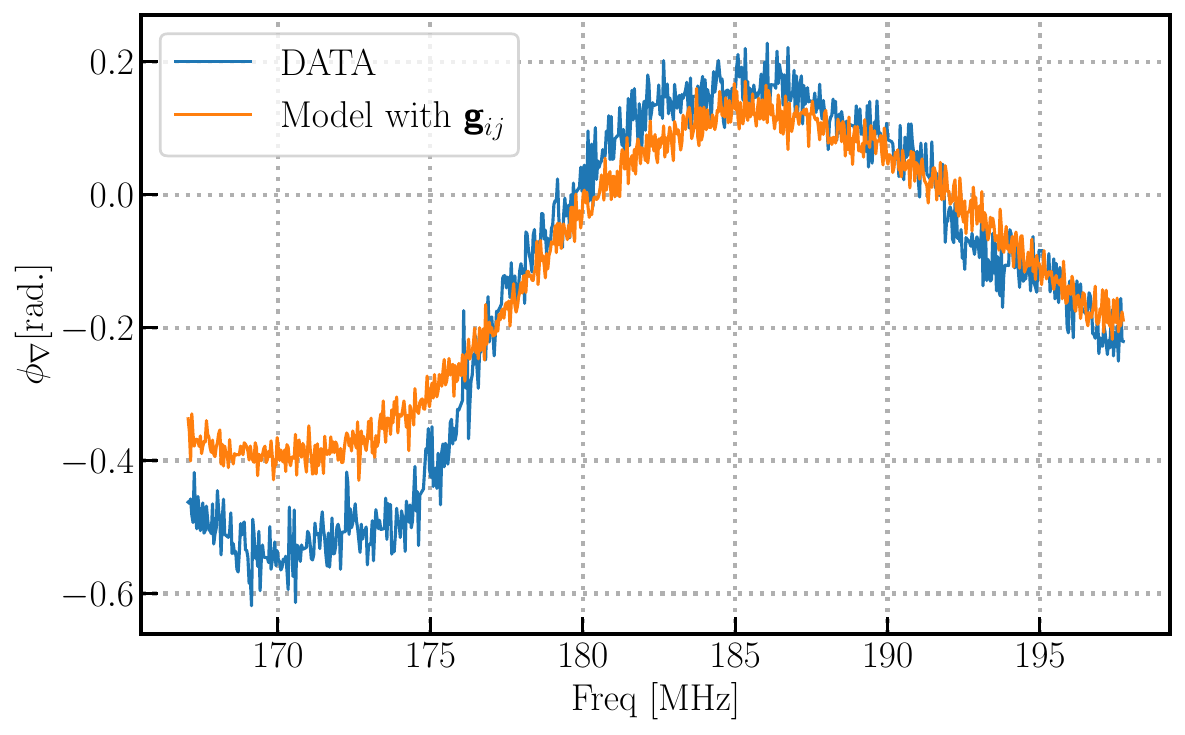}
    \caption{Comparing closure phase spectrum Data (\textit{blue}) and Model with baseline-dependent gains (\textit{orange} line)  for EoR1, 28~m baseline length.}
    \label{fig: bphase_with_bl}
\end{figure}

The eq. \ref{eq:v_ij} modifies to $V_{ij}^{\rm mod}(\nu) =  V_{ij}^{\rm m}(\nu)\mathbf{g}_{ij}(\nu);$ in scenarios incorporating baseline-dependent gains, where $\mathbf{g}_{ij}(\nu)$ represents the baseline-dependent gain factor. We introduced the baseline-dependent gains using a simple uniform distribution in the $\mathbf{g}_{ij}(\nu)$ phase with unity amplitude. The scaling factor introduced in the $\mathbf{g}_{ij}(\nu)$ sampling is set to approximately match the RMS phase of the binned averaged closure phase of the DATA. We chose brute force method to find the optimal scaling factor for a given EoR field, with a the single scaling factor for given EoR field. Figure \ref{fig: bphase_with_bl} shows the binned averaged closure phase of DATA and Model with $\mathbf{g}_{ij}$. The contribution due to the baseline dependent gains on the binned averaged closure phase about $0.05$ rad in both EoR0, EoR1 fields, which is the RMS of the ratio between the Model with $\mathbf{g}_{ij}$ and Model closure phases.
From here onwards, we use two variants of models in the analysis, the first is a forward Model without baseline dependent gains and the second is a Model with $\mathbf{g}_{ij}$.

\section{Data processing} \label{section: data_processing}
The following section provides the basic data processing steps we incorporated into this analysis. The complete schematic flowchart of the data structure is shown in \ref{fig: data_flowchart}.

\begin{figure*}
    \centering
    \includegraphics[scale=0.45]{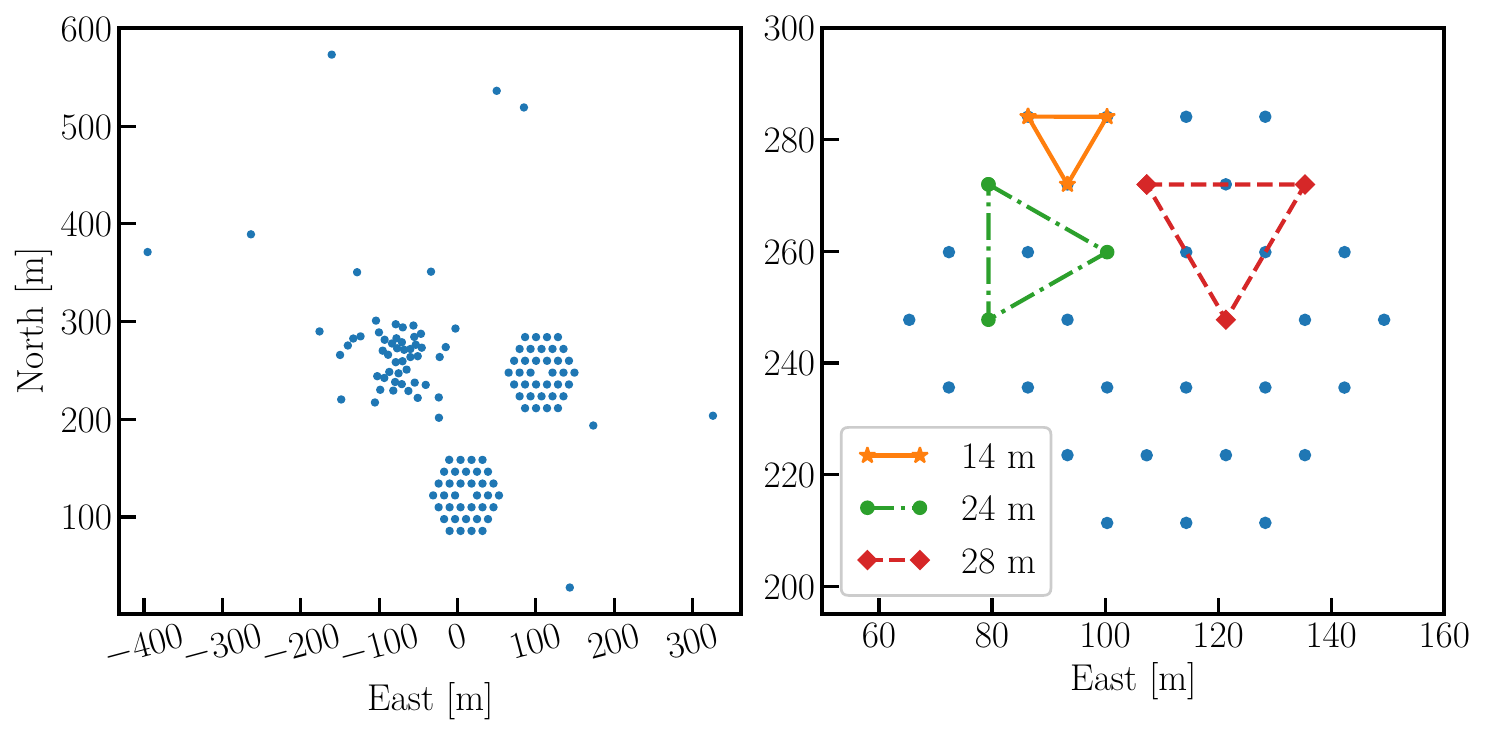}
    \caption{Left: MWA-phase II compact configuration. Right: Northern hexagon showing four equilateral triad configurations, 14 metres \textit{(orange stars)}, 24 metres \textit{(green circles)}, and 28 metres \textit{(red diamonds)}.}
    \label{fig: MWAII_config}
\end{figure*}
\subsection{DATA binning} \label{subsection: data_structure}
In the first step, the repeated night-to-night observations are sorted based on the Local Sidereal Time (LST) and Julian Date (JD); see figure \ref{fig: obs_info}. We determined the 14~m, 24~m, and 28~m redundant baseline triads from both Hexagonal configurations of MWA (see an example in figure~\ref{fig: MWAII_config} (right panel) for which the visibilities are estimated). A given triad \{a, b, c\} includes $N_{\rm vis} = 3$ visibilities which correspond to $\{V_{\rm ab}, V_{\rm bc}, V_{\rm ca}\}$.  The number of triads $(N_{\rm triads})$ varies depending on the baseline length. In our case, the $N_{\rm triads}$ are 47, 32, 29 for 14~m, 24~m, and 28~m baselines, respectively. Please note that, when accurately measured, the 24~m baseline is $14\sqrt(3)\approx24.25$ m; however, we chose the former for the simple denomination. On the other hand, the 14~m and 28~m baselines are nearly accurate for the antenna positional tolerance of MWA tiles. Each dual-polarisation observation was made for 112 seconds, which included $N_{\rm timestamps}=14$ each with 8 seconds of averaged data and a frequency resolution of 40 kHz, which provides a total of $N_{\rm channels}=768$ frequency channels with a bandwidth of 30.72 MHz. The entire observations can be restructured into;
\begin{multline} \label{eq: Nobs}
   N_{\rm obs} \equiv  \{N_{\rm LST} , N_{\rm JD}, N_{\rm timestamps}, \\ 
   N_{\rm pol}, N_{\rm triads}, N_{\rm vis}, N_{\rm channels}\} 
\end{multline}

\subsection{RFI flagging} \label{subsection: RFI Flagging}

The MWA high band ($167-197$ MHz) lies in the digital television (DTV) broadcasting band; thus, we expect RFI to be present in our dataset  \citep{Offringa_2015}, which in some cases can completely dominate the useful data from the observations. As mentioned in the previous sections, since our analysis required keeping all the frequency channels from our datasets, we did not perform any frequency channel-based RFI flagging in the data preprocessing step. Instead, we incorporated \texttt{SSINS} \citep{Wilensky_2019}, which is designed to detect faint RFI in the MWA data, to either discard the entire frequency band or keep it based on the RFI occupancy of the dataset. Instead of assuming a persistent RFI along a frequency channel, we check for the RFI along the observation time (i.e., along the $N_{\rm timestamps}$ axis).

\begin{figure*}
    \centering
    \includegraphics[scale=0.5]{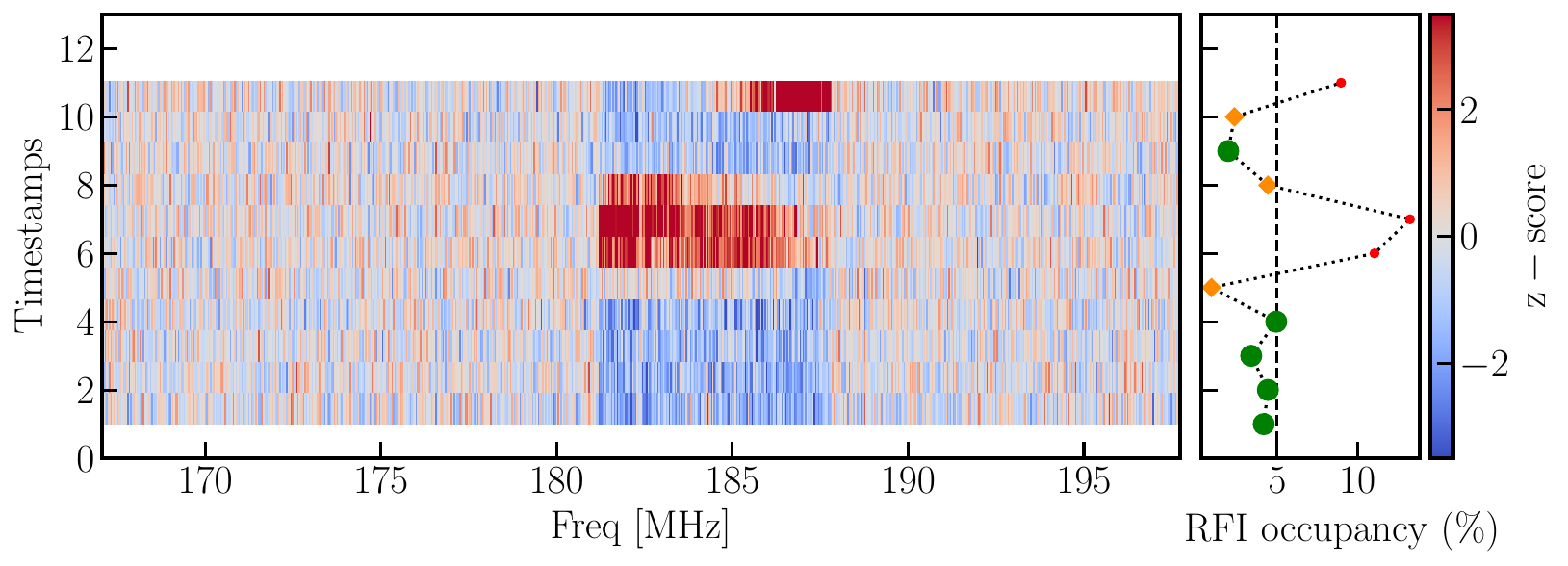}
    \caption{\textit{Left:} SSINS $z$-score for a visibly RFI-affected obsID for XX-polarisation and cross baselines. The first and last timestamps are avoided in the analysis. \textit{Right:} A $z-{\rm score}$ threshold of 2.5 was chosen to identify RFI-affected channels and timestamps, then at the first iteration of RFI flagging (whole timestamp) was chosen based on the RFI occupancy of 5\%. The timestamps corresponding to the \textit{red} points were completely discarded in the first step, and the rest of the \textit{orange} diamonds were considered good. \textit{Right:} Since the two adjacent timestamps $(\rm T_{\{i+1\}}, T_i)$ is used to estimate the $z$-score, additional timestamps were flagged in the second iteration to finally get the good timestamps shown with \textit{green} dots.}
    \label{fig: SSINS-occupancy-single}
\end{figure*}

\begin{figure*}
    \centering
    \includegraphics[scale=0.45]{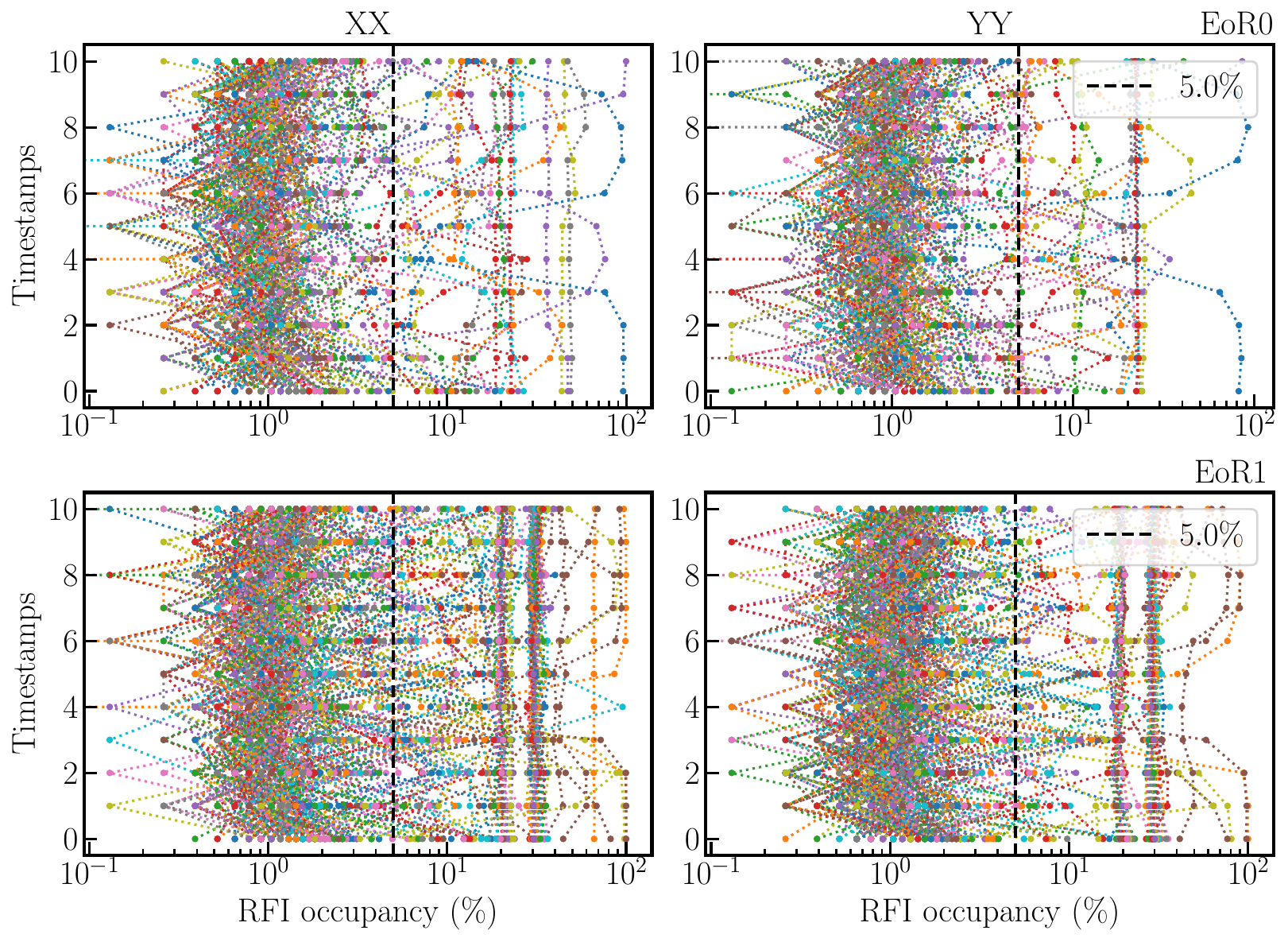}
    \caption{RFI-occupancy levels of the full dataset used in this work were obtained using SSINS. For a $z$-score threshold of 2.5, the majority of the dataset (EoR0: XX-82.5\%, YY-82.8\%; EoR1: XX-78.8\%, YY-80.9\%) shows RFI occupancy < 5\%. }
    \label{fig: SSINS-occupancy-all-data}
\end{figure*}

The flagging was performed based on the RFI $z$-score of an observation. Note that the $z$-score was estimated at successive adjacent timestamps to measure if any faint or persistent RFI was present in the data across all timestamps (see figure  \ref{fig: SSINS-occupancy-single}). We took a $z$-score threshold of 2.5, below which the data was considered good, and the channels where the $z$-score exceeded the threshold were considered RFI-affected. Then, we independently estimated the level of such RFI-affected channels along the frequencies at each timestamp and checked if the RFI occupancy at a given timestamp was more than $5\%$.  
As a first step in selecting good timestamps, we chose an RFI occupancy level of $5\%$ as a threshold. We discarded the entire timestamp if the RFI occupancy exceeded this threshold. Figure \ref{fig: SSINS-occupancy-all-data} presents RFI occupancy for the entire dataset. Since \texttt{SSINS} $z$-scores are estimated relative to the successive adjacent timestamps, it might be difficult to quantify whether the RFI leakage between the adjacent timestamps (where one is good and another is bad) is there or not. Therefore, in the second step, we again flagged all the timestamps based on whether they had a bad neighboring timestamp. The flagging provides a masked array, further propagated through other data processing steps.
\subsection{Triad Filtering}

The presence of faulty tiles or dipoles/antennae corrupts the voltages recorded by the correlator; therefore, we are required to cross-check the visibilities at each antenna triad. The easiest way was to perform a geometric median-based rectification on the closure phase. We performed a two-step median rectification on the closure phase. 
For a given observation, the data structure of the triad filtering can be shown as follows:
\[ \phi_\nabla \equiv \{N_{\rm pol}, N_{\rm triads}, N_{\rm timestamps}, N_{\rm channels}\}\]
First, we estimated the median absolute deviation (MAD = ${\rm Median} (|X_i - \bar{X}|)$) of the closure phases against the mean along the $N_{\rm triads}$,
\[ \phi_\nabla^{\rm MAD} \equiv \{N_{\rm pol}, N_{\rm triads}, N_{\rm timestamps}, N_{\rm channels}\}\]
and then we estimated the mean of the MAD (i.e. $\rm MAD_{\rm mean}$) along the $N_{\rm channels}$ axis. Finally, we estimated the MAD of the  $\rm MAD_{\rm mean}$.
\[ \mu\{\phi_\nabla^{\rm MAD}\} \equiv \{N_{\rm pol}, N_{\rm triads}, N_{\rm timestamps}, 1\}\]
This step provides a single value of the $\rm MAD_{\rm mean}$ for a given triad at every timestamp. 
\[ {\rm MAD}\{\ \mu\{\phi_\nabla^{\rm MAD}\}\} \equiv \{N_{\rm pol}, N_{\rm triads}, N_{\rm timestamps}, 1\}\]
Finally, we masked the triads if the mean of the MAD is greater than 1$\sigma \approx 1.4826$ of MAD and considered the triad performing poorly at that given timestamp.

\subsection{Coherent Time Averaging} \label{subsection: coherent_averaging_time}
The coherent averaging gives us an estimate of a timescale up to which the sky signal can be assumed identical and averaged coherently to improve sensitivity. It can be estimated by measuring the variation in the sky signal with time for a fixed pointing. Indeed, these vary with instrument and frequency of observation since the beam sizes are different. To check this with MWA, we used a continuous drifted sky simulation (FG and H{\scriptsize{I}}) under ideal observing settings (i.e., unity antenna gains, equal antenna element elevation from the ground) for $\approx 0.5$ hours while keeping a fixed zenith pointing. The sky moves about $\approx 7.5^\circ$ in $0.5$ hours, which is less than the MWA beam size of $\approx 9^\circ-7.5^\circ$ at the shortest ($14$~m) triad, thus justifying the simulation time range of $0.5$ hour. 

\begin{figure}
    \centering
    \includegraphics[scale=0.5]{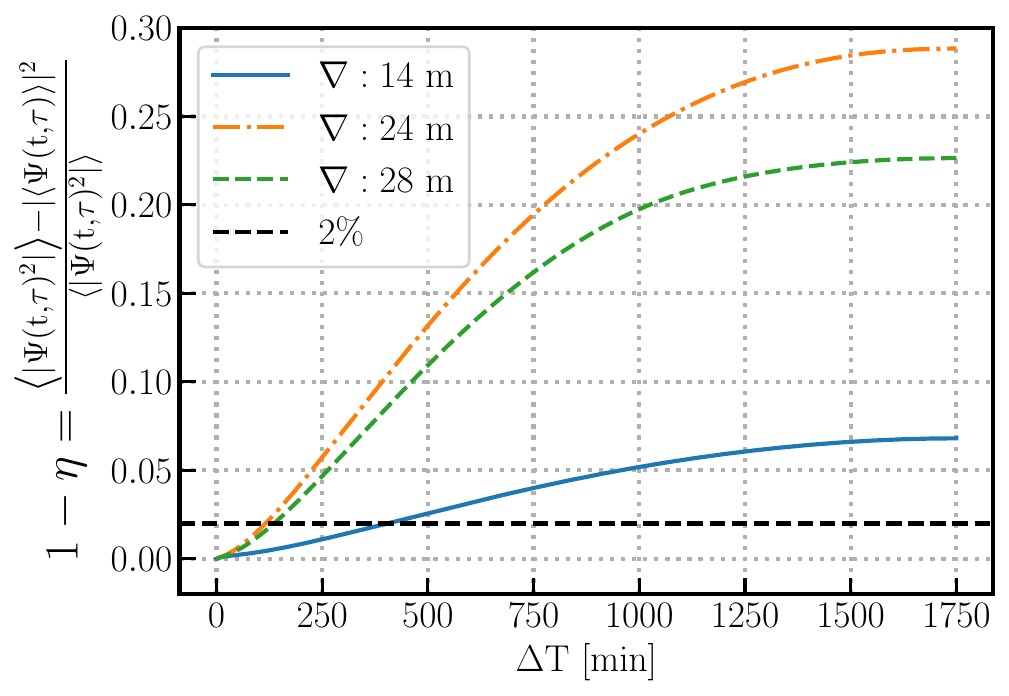}
    \caption{Fractional power loss for different triad configurations (shown by \textit{coloured} lines). The 2\% threshold level is shown with the \textit{black} dashed horizontal line.
    The intersection between the threshold line and different triad loss provides an estimate of the time the data can be averaged coherently along LST.
    The coherent averaging time for the 14~m, 24~m, and 28~m baseline triads is roughly 408, 130, and 120 seconds, respectively.}
    \label{fig: coherence-time-all-cases}
\end{figure}

We simply added the ideally simulated visibilities (FG and H{\scriptsize{I}}) and estimated the closure phase power spectrum as a function of time for higher delay $(|\tau|> 2~\mu \rm s)$. We used a fractional signal loss of $2\%$ to measure the coherence threshold, a similar approach used by \citet{Keller_2023}.
The fractional loss in power is defined as,
\begin{equation}
    1-\eta = \frac{\left<|\psi_\nabla(t, \tau)^2|\right> - |\left<\psi_\nabla(t, \tau)\right>|^2}{\left<|\psi_\nabla(t, \tau)^2|\right>}
\end{equation}
The choice of $|\tau|> 2~\mu \rm s$ is to choose the timescale based on the loss of H{\scriptsize{I}}~signal, which is where it would be significant, namely, the higher delay modes. Figure \ref{fig: coherence-time-all-cases} shows the fractional loss of coherent H{\scriptsize{I}}~signal power as a function of averaging timescale for three triad configurations. We found the coherence averaging time for $\{\nabla: 14, 24, 28\}$ triads to be approximately 408, 130, and 120 seconds, respectively. 

\section{Results} \label{section: results}

\begin{figure}
    \centering
    \includegraphics[scale=0.40]{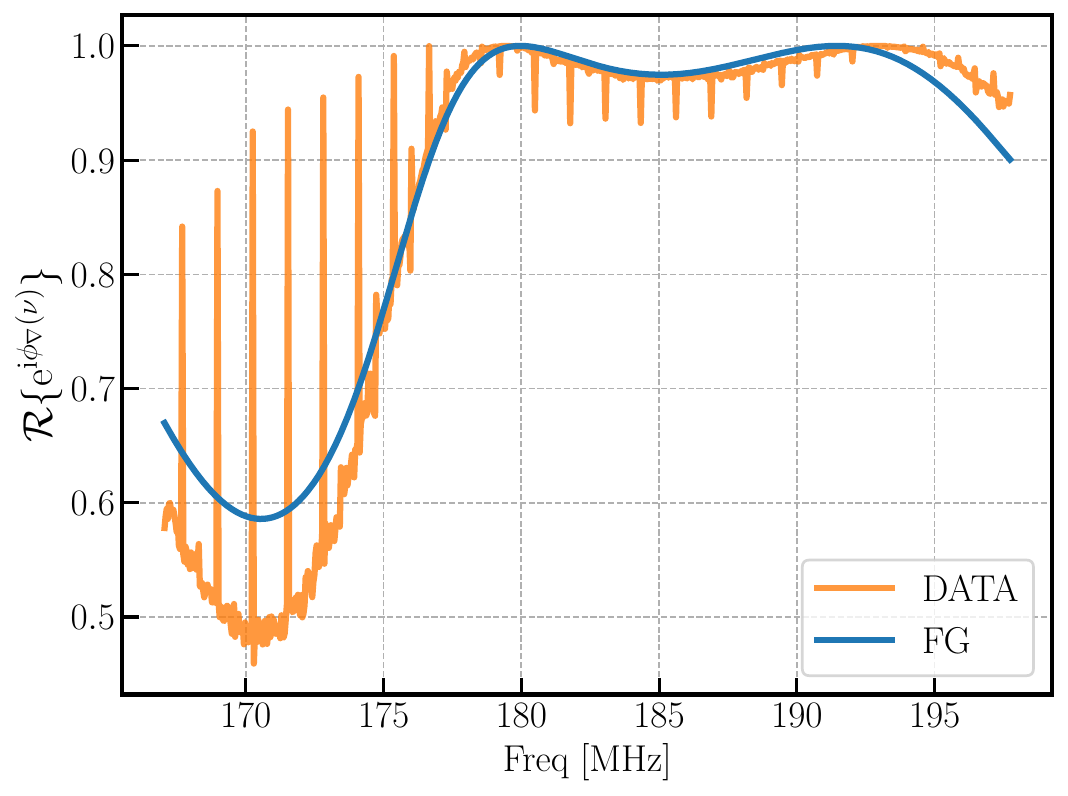}
    \caption{The real part of the complex exponent of the closure phase for XX polarisation 28~m baseline. The data matches with the foreground simulations. For the sanity check, the foreground simulation of the same is plotted over the data. Despite eliminating element-based bandpass gains, the data contains periodic spikes corresponding to the 1.28~MHz coarse channel edges of the MWA bandpass, indicating possible systematics of baseline-dependent origin.}
    \label{fig: closure_bandpass}
\end{figure}

The bandpass structure of the MWA leaves significant systematics in the edge channels. Usually, these edge channels get flagged in the early data preprocessing step. However, we did not apply the flags since we require the full observing band for the delay spectrum analysis. The closure phase must be free of phase errors associated with the individual antenna elements. Thus, we expected the element-based bandpass structure to be removed in the closure phase. However, we still observed some residual bandpass structures in the closure phases; see figure \ref{fig: closure_bandpass}. It can happen if there are some baseline-dependent gains present in the data in addition to antenna-dependent gains. We validated our hypothesis by developing a simple bandpass simulator, where we introduced an additional bandpass structure to MWA data that consisted of both element-based and baseline-dependent terms. The bandpass persists if baseline-dependent gains are present in the data, which otherwise gets completely removed if only antenna-based gains are present in the data (see appendix figure  \ref{fig: bandpass-sim}). In total, about 128 frequency channels are affected by the bandpass, which accounts for about 16\% of the total bandwidth. 

\subsection{Mitigation of baseline-dependent bandpass effects}

We investigated two approaches to overcome the presence of baseline-dependent edge channel effects, the first being the Non-uniform Fast Fourier transform (NFFT), where we tried to avoid the bandpass-affected channels in the Fourier transform. The second Gaussian Process Regression (GPR) based data-inpainting to estimate the missing channel information at the location of the spikes in the bandpass spectrum.

\subsubsection{Gaussian Process Regression (GPR)} \label{subsubsection: GPR}

GPR \citep{Wiener_1949, Rybicki_1992, Robertson_2015} is a popular supervised machine learning method. GPR has previously been used in foreground subtraction \citep{Mertens_2018, Ghosh_2020}, and data inpainting \citep{Trott_2020, Kern_2021} and characterisation \citep{Pagano_2023} for 21-cm cosmology studies.
We follow a similar formalism to \cite{Kern_2021}. In the first step, we identified all the bad edge channels in the bandpass and flagged them in the closure phase. As mentioned before, edge channel contamination is present in the MWA data. The data at 40 kHz resolution has 768 frequency channels and 128 contaminated edge channels. We implemented GPR on the complex exponent of the closure phase ${\rm exp}(i\phi_\nabla (\nu))$, with the real and imaginary components separately. The GPR implementation was rather simplistic since the complex phase varies between $[0, 1]$. We used the Matérn kernel as model covariance in our analysis. To optimise the kernel hyperparameters, we used the \texttt{scipy}-based \texttt{L-BFGS} \citep{liu1989limited} to find the minima of the objective function. The optimisation was reiterated over ten times to ensure the kernel hyperparameters' convergence. 
Note that for a given frequency range, we applied GPR to the entire $N_{\rm obs}$ (see eq. \ref{eq: Nobs}) separately; therefore, the kernel Hyperparameters are also different for each closure phase frequency spectrum. 

\begin{figure}
    \centering
    \includegraphics[scale=0.40]{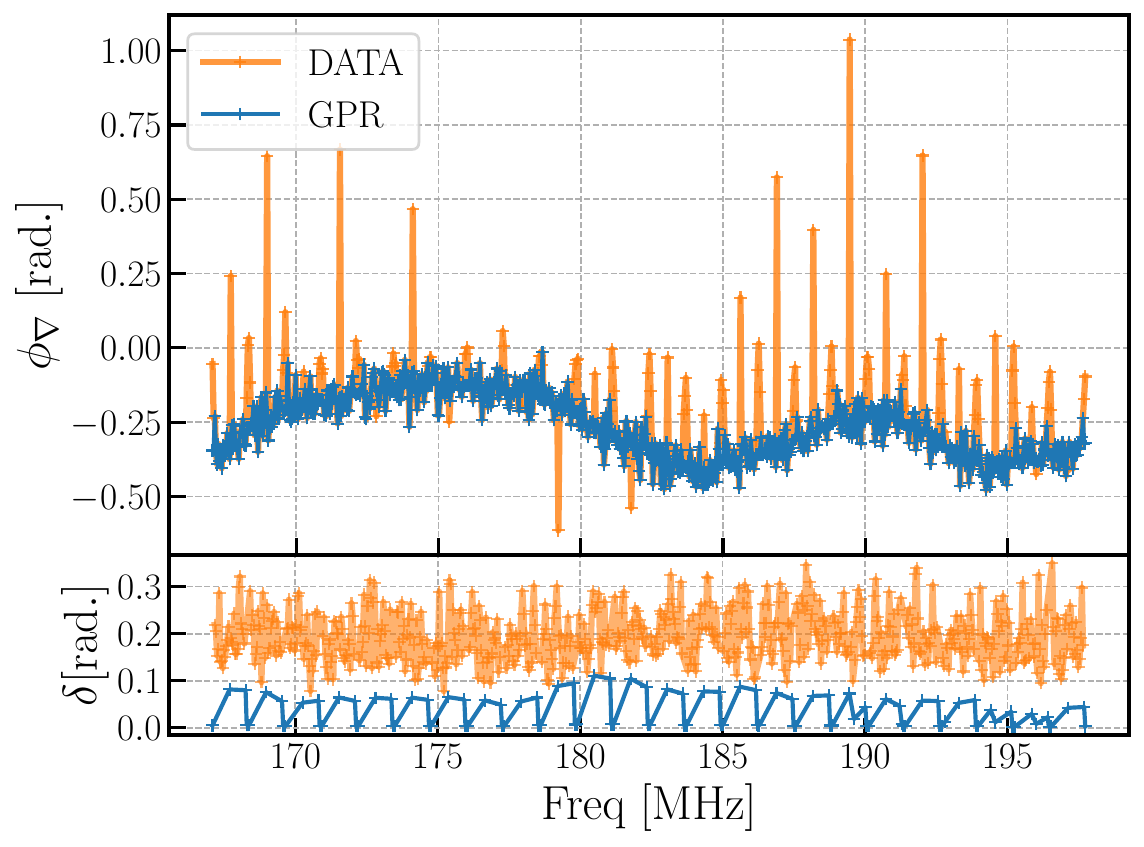}
    \caption{Top: bin-averaged closure phase of the data shown by \textit{blue} and GPR reconstruction \textit{orange} line. Bottom: RMS of the difference of closure phase. The edge channels were removed while calculating the RMS in the data, whereas only the edge channels were retained for the GPR.}
    \label{fig: GPR_recon}
\end{figure}

Figure~\ref{fig: GPR_recon} shows the closure phase of the data and GPR reconstruction. In the top panel, the data can be seen with spikes at regularly spaced edge channels of $\approx 1.28$ MHz intervals. The interpolated values of the closure phase are plotted over the data. The bottom panel shows the difference in the RMS in the closure phase along the frequency axis. Note that, while estimating the relative difference in the data closure phase, we avoided the noisy edge channels, whereas, for the GPR case, we only included the relative difference near the edge channels. This will enable us to query whether the GPR closure phase has a similar variation across the frequency compared to the data. It can be seen that the RMS of the data is higher than the GPR values, which means that the GPR has performed quite well. We used the Python-based module \texttt{GPy}\footnote{\url{https://github.com/SheffieldML/GPy}}  for the GPR implementation. 

\subsubsection{Non-uniform Fast Fourier transform (NFFT)} \label{subsubsection: NFFT}

NFFT is a well-known method to get the Fourier transform of the data with missing samples \citep{Dutt_1993, Beylkin1995}. To estimate the FFT of bandpass-affected data, first, we removed the 128 edge channels from the data and estimated the $\Psi_\nabla(\nu)$ (a Fourier conjugate of eq.\ref{eq:psi_tau}), which is then supplied to the NFFT function to get $\Psi_\nabla(\tau)$. We used a Python-based \texttt{NFFT} \footnote{\url{https://github.com/jakevdp/nfft}} package to develop the NFFT functions. 

\begin{figure}
    \centering
    \includegraphics[scale=0.40]{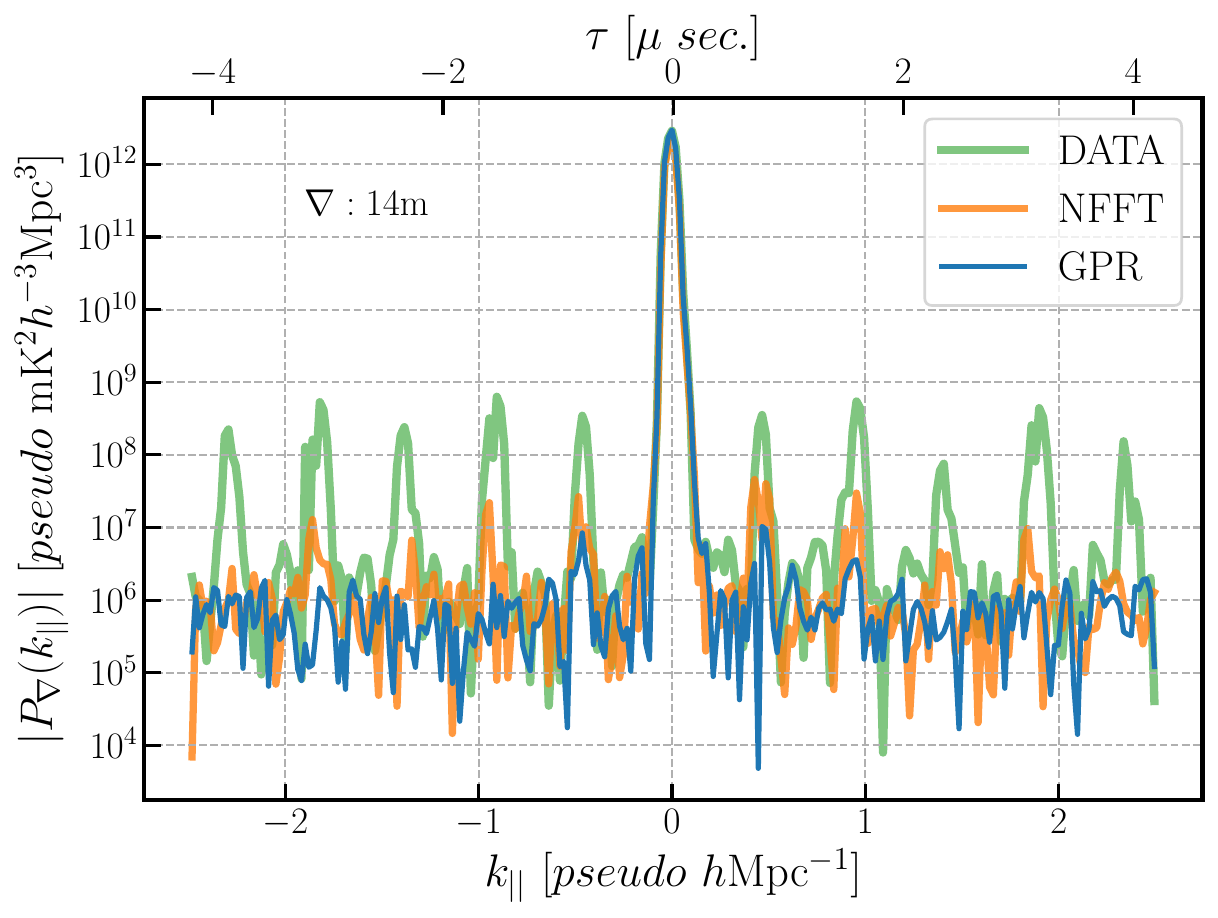}
    \caption{Absolute cross-power spectra of the DATA, NFFT, and GPR are shown with \textit{colored} lines. MWA DATA suffers from baseline-dependent bandpass structure (see the regularly spaced spikes, which correspond to $\approx$ 1.28 MHz). NFFT significantly dampens the bandpass, but the spikes persist in the spectrum. GPR reconstruction shown by the \textit{blue} line provides the cleanest spectra.}
    \label{fig: abs-power_comparison}
\end{figure}

The absolute values of the closure phase delay cross-power spectrum are shown in figure \ref{fig: abs-power_comparison}. It can be seen that the data is highly affected by the excess systematics in the power, evident by the periodic spikes. NFFT significantly reduces the bandpass systematics but does not eliminate it entirely. Since the GPR performed best between the two methods, we adopted only the GPR for the later analysis. From now on, we will be using `GPR-reconstructed DATA' as `DATA' for the entirety of the paper.

\subsection{Cross power spectrum estimation} \label{subsection: estimating cross-power}

We proceeded to the power spectrum estimation after eliminating a significant part of the baseline-dependent bandpass contamination using GPR inpainting. Based on the LST-JD of the observations (see figure. \ref{fig: obs_info}, we split the dataset equally along the JD axis (i.e., $N_{\rm JD}=2$). 
We binned the data along LST according to the coherent averaging time of MWA, which we already estimated for the redundant 14~m, 24~m, and 28~m baselines, resulting in 5, 14, and 17 LST bins, respectively.
The first level of weighted averaging (i.e., coherently averaging) was done to the bispectrum (visibility triple product) and effective foreground visibilities, $(V_{\rm eff})$, that lie within the same LST bin. The weights are the number of good observations left after being rectified by the RFI flagging and triad filtering within the given LST bin for a given polarisation and triad. 

In the next step, we estimated the delay spectra, $\Psi_{\nabla}(\tau)$, from the phase of the LST binned averaged closure spectra at each LST bin. It resulted in delayed spectra data structure, 
\[\{ N_{\rm LST}, N_{\rm JD}, N_{\rm pol}, N_{\rm triads}, N_{\rm delays}\}; N_{\rm delays}=N_{\rm channels}\]
Finally, we estimated the cross-power between the unique triad pairs of the delay spectra across the two JD-bins according to Equation~(\ref{eq: power_kpar})
where,
\begin{equation}
    \tilde \Psi_{\nabla}(\tau) . {\overline{\tilde \Psi_{\nabla}^\prime(\tau)}} = \frac{1}{\prescript{N_{\rm triads}}{}{C}_{2}}\sum_{i,j}^{N_{\rm triads}} \tilde \Psi_{\nabla}(i, \tau) . {\overline{\tilde \Psi_{\nabla}(j, \tau)}}, \quad i>j,
\end{equation}
where $i, j$ runs over $N_{\rm triads}$ (upper triangle of the $\{i,j\}$ pairs) from the first and second JD bins. The normalising factor of 2 arises since phases only capture half the power in the fluctuations. After this operation, the data structure of the binned averaged cross $P_\nabla(k_{||})$ becomes 
\[\{N_{\rm LST}, \prescript{N_{\rm JD}}{}{C}_{2}, N_{\rm pol}, \prescript{N_{\rm triads}}{}{C}_{2}, N_{\rm delays}\}; \prescript{N_{\rm JD}}{}{C}_{2}=1.\]

We took the weighted mean (i.e., incoherently averaged) along the triad pairs, where the weights were propagated from the previous step (refer to data flowchart~\ref{fig: data_flowchart} for details).
As the sky varies with LST, we applied the inverse variance weights along the LST axis and averaged them to get the final estimates of the power spectrum. The same operation was done for the imaginary part of the data to get an estimate of the level of systematics in the power spectrum.

Ultimately, we incoherently averaged the two polarisations and downsampled the original delays to the effective bandwidth of the applied window function (i.e., $\approx 12.9$ MHz). We used \texttt{Scipy}-based \texttt{BSpline} to interpolate at the new downsampled delays. The final estimates of the closure phase power spectra are shown in figure \ref{fig: cross-power-EoR0} and \ref{fig: cross-power-EoR1} for the EoR0 and EoR1 fields, respectively. 
 
\subsection{Error estimation} \label{subsection: error_estimation}

The uncertainties on the power spectrum can be estimated in multiple ways. We primarily focused on estimating the uncertainties in two ways, the first being `noise+systematics' and the second only noise, where we tried to mitigate the systematics. 

\subsubsection{Noise+systematics} \label{subsection: noise_sys_error}

To estimate uncertainties in power, we increased the number of samples that go into the uncertainty estimation by splitting the JD-axis into four parts (i.e., $N_{\rm JD} = 4$). This led to the data $(\Psi_\nabla(\tau))$ structure  being $\{N_{\rm LST}, N_{\rm JD}=4, N_{\rm pol}, N_{\rm triads}, N_{\rm delays}\}$. 
We similarly estimated the cross-power of the $N_{\rm triads}$ along the unique pairs of $N_{\rm JD}$ axis. This operation provided us with $\{N_{\rm pol}, N_{\rm LST}, \prescript{N_{\rm JD}}{}{C}_{2}=6, \prescript{N_{\rm triads}}{}{C}_{2}, N_{\rm delays}\}$ unique power spectra. The weighted mean power was estimated by along $\prescript{N_{\rm triads}}{}{C}_{2}$ where the weights are coming from the number of good observations that went into the $\Psi_\nabla$ for a given triad.

Next, We estimated the weighted variance on the power using the standard error of the weighted mean provided in \cite{cochran1977},
\begin{multline}\label{eq: wtd_var_}
    ({\rm SEM}_{\rm wtd})^2 = \frac{n}{(n-1)(\sum w_i)^2} [ \sum (w_i X_i - \bar{w}  \bar{X}_{\rm wtd})^2 \\
    - 2\bar{X}_{\rm wtd}\sum (w_i -\bar{w}) (w_i X_i - \bar{w}  \bar{X}_{\rm wtd})
    + \bar{X}_{\rm wtd}^2 \sum (w_i -\bar{w})^2] \, ,
\end{multline}
where $w_i$ are the weights, and $n$ represents the weight count. Note that since the $N_{\rm JD} = 4$, we are required to normalise the variance. Figure~\ref{fig: data_flowchart} illustrates the detailed data structure flow for the noise estimation.

\begin{figure*}
    \centering
    \includegraphics[scale=0.35]{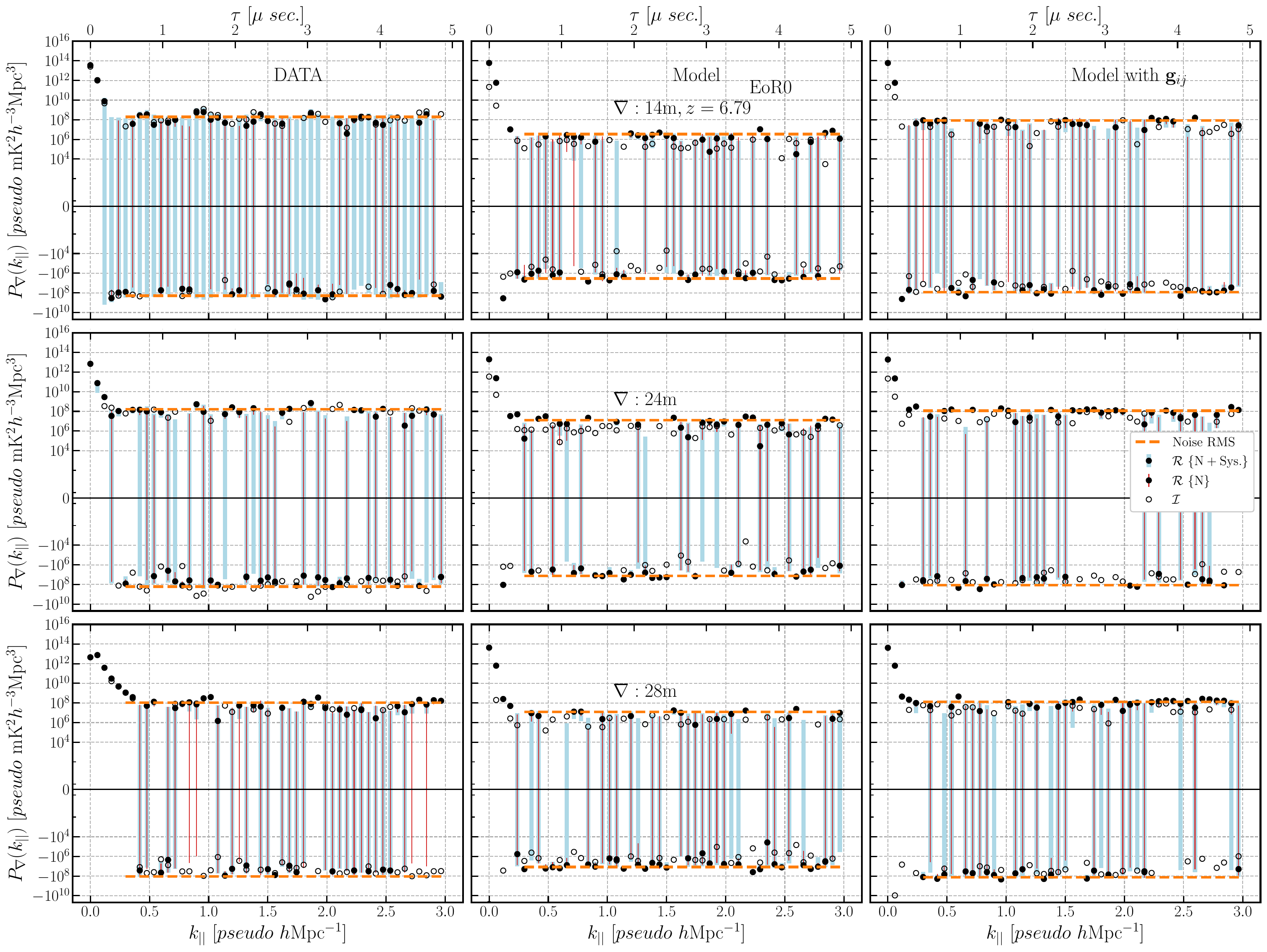}
    \caption{Cross power spectrum of the closure phase delay spectrum for EoR0 observing field. The left panel represents the DATA, middle panel Model \{FG + H{\scriptsize{I}} + noise\} and the right Model with $\mathbf{g}_{ij}$. The top, middle, and bottom panels show the power spectra for 14~m, 24~m, and 28~m baseline lengths, respectively. The real part (\textit{filled circles}) denotes the power, while the imaginary (\textit{hollow circles}) represents the systematic level in the data. 2$\sigma$ uncertainties are shown for two scenarios; the noise+Systemtatics are shown with \textit{skyblue} error bars while noise-only is shown with \textit{red} error bars. The RMS level for $k_{||}\geq 0.15~ pseudo~ h{\rm Mpc^{-1}}$ is shown using \textit{orange} dashed line. The noise+Systematics uncertainties are >10 times the noise-only uncertainties at low delays ($k_{||}<0.5~pseudo~h{\rm Mpc^{-1}}$), while fluctuating between > 4-8 times at higher delays.}
    \label{fig: cross-power-EoR0}
\end{figure*}

\begin{figure*}
    \centering
    \includegraphics[scale=0.35]{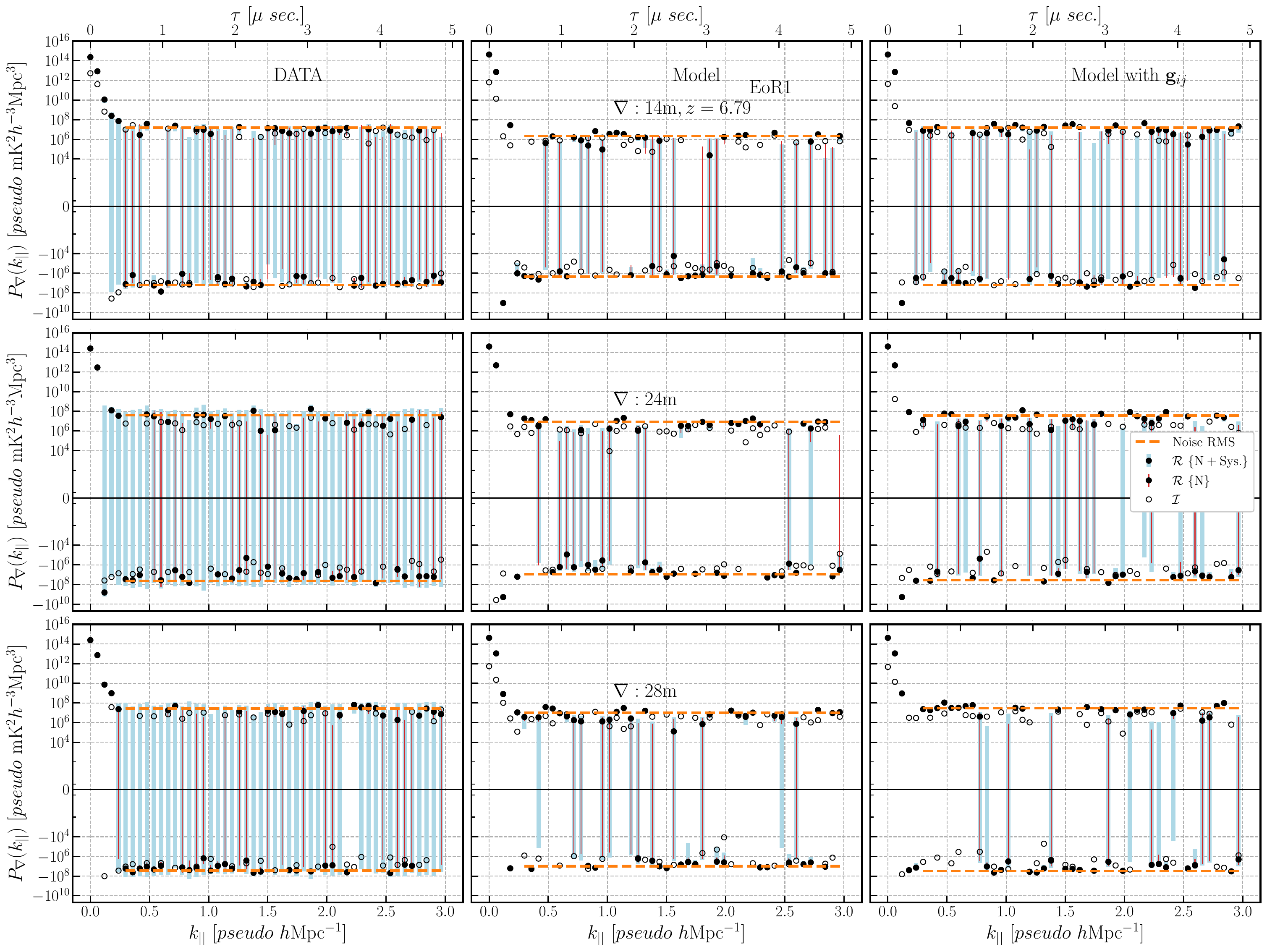}
    \caption{Same as Fig.\ref{fig: cross-power-EoR0} but for EoR1 observing field.}
    \label{fig: cross-power-EoR1}
\end{figure*}

\subsubsection{Noise} \label{subsection: noise_error}

The uncertainties for the only noise case follow a similar procedure as the previous one, with the same JD split (i.e., $N_{\rm JD} = 4$), the cross-power is estimated, which led to the data structure $\{N_{\rm pol}, N_{\rm LST}, \prescript{N_{\rm JD}}{}{C}_{2}=6, \prescript{N_{\rm triads}}{}{C}_{2}, N_{\rm delays}\}$. 
After this operation, we took the difference in the power spectra between the unique pairs of JD (i.e., the same sky signal). 
The only noise scenario can be understood assuming the cross-power between the two independent JD-bins correlates with the common signal and systematics across the triads. Assuming the power is coherent within the LST bin, the successive unique difference in the power within the LST bin eliminates the correlated power and systematics, leaving only the noise-like residuals. Also, since the differences eliminate the sky signal, the LST variation in the difference power is minimal; thus, we collapsed our datasets into a single axis and estimated the weighted standard deviation of the differenced power to get the final noise-like uncertainties. Finally, again, we took the weighted variance using eq.~\ref{eq: wtd_var_}.



\subsection{Validation} \label{subsection: validation}
We performed a two-sided KS test on the closure phase power spectra of the data and two model variants for the statistical comparison. The null hypothesis was rejected in all scenarios with the Model (without baseline dependent gains); however, it failed to reject the null-hypothesis in all scenarios when using the Model with $\textbf{g}_{ij}$. 
The former implies that the data and the model uncertainties were unlikely to be drawn from the same distribution, whereas the latter concludes the contrary. The test results are provided in table~\ref{tab: KS-test_main}. The test results for the Model are not unexpected since we can see that the RMS floor levels of the data and Model are sufficiently different, which match the data and Model with $\textbf{g}_{ij}$. 
We modelled the excess RMS in the data as arising from baseline-dependent gain factors, although it is believed to have originated from the systematics and residual RFI. Note that we do not claim that the excess noise in the data is solely due to the baseline dependent systematics; however, if the argument is true, the baseline dependent gains introduced in our analysis suffice for the excess power in the data.

\begin{table}
\centering
\caption{2-sided KS test comparison between the Data and Model, Data and Model with $\mathbf{g}_{ij}$ at $k_{||}>0.15~ [pseudo ~h\rm Mpc^{-1}]$.}
\begin{tabular}{|c|c|cc|ll|} 
    \hline
  \multicolumn{6}{|c|}{2-sided KS-test}\\ \hline
 field&$\nabla$& \multicolumn{2}{c|}{Model}& \multicolumn{2}{c|}{Model with $\mathbf{g}_{ij}$}\\ \hline
  && $p$-value & statistic& $p$-value&statistic\\ 

     & $14$~m & \fbox{$6\times 10^{-6}$}&$0.51$&  \colorbox{gray}{$0.24$}&$0.21$\\ 
EoR0 & $24$~m & \fbox{$8\times 10^{-4}$}&$0.40$& \colorbox{gray}{$0.50$}&$0.17$\\ 
     & $28$~m &  \fbox{$2\times 10^{-6}$}& $0.53$&  \colorbox{gray}{$0.09$}&$0.25$\\ \hline
          & $14$~m & \fbox{$2\times10^{-3}$}& $0.38$& \colorbox{gray}{$0.84$}&$0.13$\\ 
EoR1 & $24$~m & \fbox{$3\times 10^{-2}$}& $0.29$& \colorbox{gray}{$0.68$}&$0.15$\\ 
     & $28$~m &  \fbox{$8\times10^{-3}$}& $0.34$& \colorbox{gray}{$0.24$}&$0.21$\\ \hline
\end{tabular}\label{tab: KS-test_main}

\end{table}

\subsection{21-cm power spectrum}
\begin{figure*}
    \centering
    \includegraphics[scale=0.42]{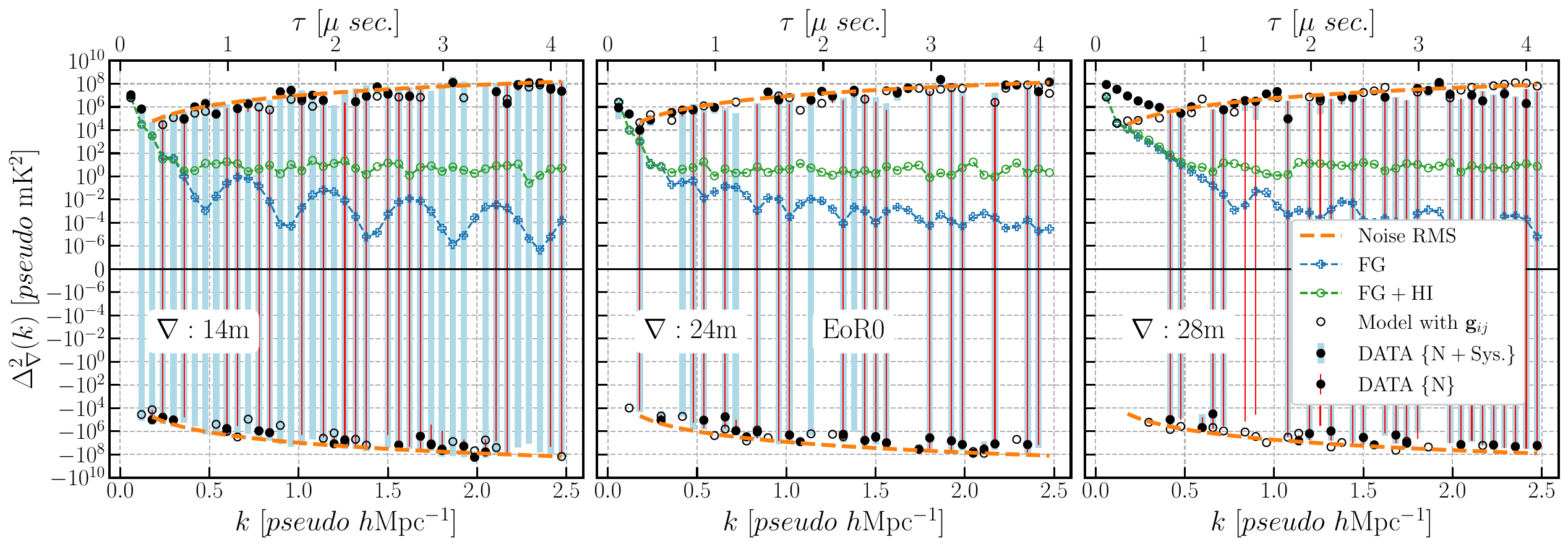}
    \caption{21-cm power spectrum from 14~m (left), 24~m (middle), and 28~m equilateral triads for the EoR0 field.}
    \label{fig: 21-cm_PS_EoR0}
\end{figure*}

\begin{figure*}
    \centering
    \includegraphics[scale=0.42]{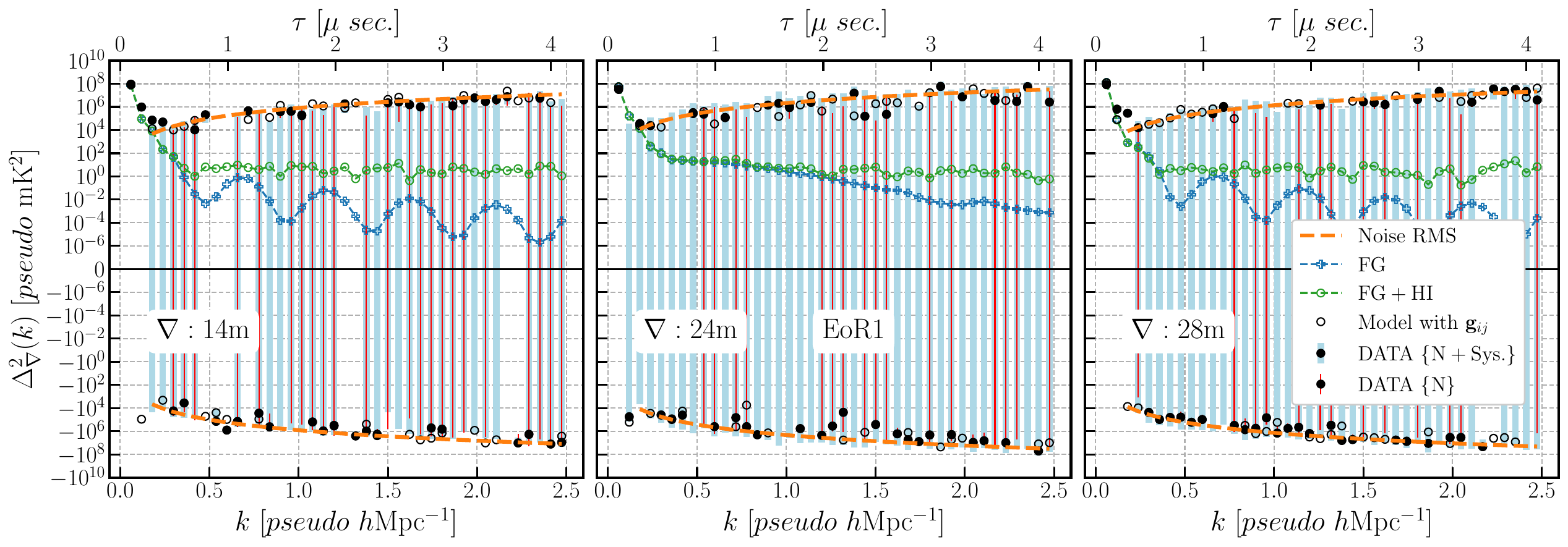}
    \caption{Same as Fig.\ref{fig: 21-cm_PS_EoR0} but for the EoR1 field.}
    \label{fig: 21-cm_PS_EoR1}
\end{figure*}

We estimated the final dimensionless 21-cm power spectrum from the closure phase power spectrum. The closure phase delay power spectrum can be written into a 21-cm power spectrum (``$pseudo$'') as follows:
\begin{equation}
    \Delta_{\nabla}^2(k) = \frac{k^3P_\nabla(k_{||})}{2\pi^2} [pseudo~ {\rm mK^2}]
\end{equation}
where, $k^2 = k_\perp^2 + k_{||}^2$,  with $k_\perp = \frac{2\pi |b_\nabla|}{\lambda D}$, where $b_\nabla$ is the baseline length of the triad, and $D$ is the cosmological comoving distance. Note that the 21-cm power spectrum estimates from the closure phase power spectrum should not be interpreted as true but rather the approximate representation of the actual 21-cm power spectrum \citep{Nithya_20b, Keller_2023}. The power spectra converted to cosmological units for EoR0 and EoR1 fields are shown in Figures~\ref{fig: 21-cm_PS_EoR0} and \ref{fig: 21-cm_PS_EoR1}, respectively.  

\begin{table*}
\caption{2$\sigma$ upper limits on 21-cm power spectrum [$pseudo~ {\rm mK^2}$]. The two estimates correspond to only-noise and noise+Systematics case.}\label{tab: 21-ps}
\resizebox{18cm}{!}{
\begin{threeparttable}

\begin{tabular}{||l||c|c|c|c|c|c|c|c|c|c|c|c|} \hline 
  \multicolumn{13}{|c|}{$\Delta_{\nabla \rm ~UL}^2 [pseudo~ {\rm mK^2}]$}\\ 
\hline
Field
   &\multicolumn{6}{c}{EoR0}&  \multicolumn{6}{|c|}{EoR1}\\ \hline
Baseline
&\multicolumn{2}{c|}{$\nabla: 14~\rm m$} 
&\multicolumn{2}{c|}{$\nabla: 24~\rm m$}
&\multicolumn{2}{c|}{$\nabla: 28~\rm m$}
&\multicolumn{2}{c|}{$\nabla: 14~\rm m$}
&\multicolumn{2}{c|}{$\nabla: 24~\rm m$}
&\multicolumn{2}{c|}{$\nabla: 28~\rm m$}\\\hline
$k$ [$pseudo~h{\rm Mpc^{-1}}$]
& N. & N.+Sys.
& N. & N.+Sys.
& N. & N.+Sys.
& N. & N.+Sys.
& N. & N.+Sys.
& N. & N.+Sys.\\
\hline\hline
$0.18$&-- & \fbox{$(392^*)^2$}&\fbox{$(188^*)^2$} & \fbox{$(207^*)^2$}&-- & --&-- & $(526)^2$&-- & \fbox{$(361)^2$}&-- & --\\
$0.24$&\fbox{$(347^*)^2$} & $(420)^2$&-- & --&-- & --&-- & $(427)^2$&-- & $(458)^2$&\fbox{$(236)^2$} & \fbox{$(314)^2$}\\
$0.30$&-- & $(534)^2$&-- & --&-- & --&$(218)^2$ & \fbox{$(263)^2$}&-- & $(503)^2$&-- & $(512)^2$\\
$0.36$&$(490)^2$ & $(608)^2$&-- & --&-- & --&\colorbox{Gray}{$(184)^2$} & $(330)^2$&-- & $(849)^2$&-- & $(572)^2$\\
$0.42$&-- & $(1562)^2$&-- & $(1065)^2$&\fbox{$(732^*)^2$} & \fbox{$(708^*)^2$}&$(474)^2$ & $(434)^2$&-- & $(1037)^2$&-- & $(762)^2$\\
$1.50$&$(5275)^2$ & $(7175)^2$&$(2466)^2$ & $(3263)^2$&$(2525)^2$ & $(3751)^2$&$(2926)^2$ & $(2748)^2$&\fbox{$(739)^2$} & $(3695)^2$&$(3758)^2$ & $(4835)^2$\\

\hline\hline

\end{tabular}
\begin{tablenotes}
   \item[a]  k-modes where the uncertainty brackets do not include zero power are masked and shown with dashes ($-$).  
   \item[b]  best limits for a given baseline triad are shown with filled Gray box and unfilled boxes.
   \item[c] limits quoted with an asterisk ($*$) might be affected by systematics or persistent residual RFI.
\end{tablenotes}
\end{threeparttable}}
\end{table*} 

Assuming the convergence to normal distribution due to the central limit theorem, we estimated $2\sigma$ [$95\%$ confidence intervals (CI)] using the uncertainties since our sample size was sufficient $(> 30)$. The upper limits on the 21-cm power spectrum [$pseudo~{\rm mK^2}$] were then estimated 
$\{ \Delta_{\nabla \rm ~UL}^2 = (\mathcal{\mu}_{\Delta^2_{\nabla}} \pm {\rm CI}) ~[pseudo~ {\rm mK^2}]\}
$
for both the EoR0 and EoR1 fields are provided in the table \ref{tab: 21-ps}, and the full table in the (\ref{tab: 21-ps_full}).

\section{Discussion} \label{section: discussion}

We used the closure phase delay spectrum technique to obtain an independent estimate of the 21-cm power spectrum for the MWA phase~II observations. These observations were centered on the EoR0 and EoR1 fields and were zenith-pointed, similar to the observing strategy of HERA. Our analysis revealed that MWA observations are possibly suffer from a baseline-dependent bandpass structure, which is especially noticeable in the edge channels. This bandpass structure results in structured bumps in the delay power spectrum (see figure \ref{fig: abs-power_comparison}), significantly contaminating the power spectrum. To address this issue, we explored two methods: Gaussian Process Regression (GPR) and Non-uniform Fast Fourier Transform (NFFT), to inpaint and mitigate the impact of the bandpass-affected edge channels on our power spectrum estimation. However, we decided against adopting the NFFT method because, although it reduced the magnitude of the bandpass, the bandpass remained evident in the NFFT delay spectra (see figure \ref{fig: abs-power_comparison}). Finally, we estimated the  21-cm power spectra using closure phase delay spectra. Additionally, we performed forward modelling in parallel with the observations to gain insights into the nature of the power spectrum under ideal observing conditions. The main findings of our analysis are summarised below.

    When we averaged closure phases across multiple timestamps within the same Local Sidereal Time (LST) bin, we noticed a significant residual bandpass structure, particularly noticeable in the edge channels (see Fig. \ref{fig: closure_bandpass}). Since closure phases are unaffected by element-based bandpass variations, we concluded that these bandpass issues cannot be simplified into element-based terms and could instead be dependent on the baseline. To test this hypothesis, we simulated the same effect on foreground visibilities (see figure \ref{fig: bandpass-sim}). We explored data-inpainting techniques to address these systematic errors to estimate closure phases in channels contaminated by baseline-dependent bandpass systematics (see figure \ref{fig: abs-power_comparison}). It is important to note that while these baseline-dependent issues are most noticeable in the edge channels, they could potentially affect all frequency channels since closure phases do not eliminate them. These issues also impact standard visibility-based power spectrum analysis methods. Understanding how the antenna layout contributes to such systematic errors is crucial for executing baseline-based mitigation strategies. Further investigation is needed to fully understand the extent to which these systematic errors affect MWA EoR power spectrum estimates. With a simple baseline dependent gains in the forward modelling, we aim to address the anomalies present in the DATA.
    
    On comparing the final closure phase power spectrum of the DATA and Model for the EoR0 field (see figure \ref{fig: cross-power-EoR0}), we found that the peak power (at $\tau=0\mu s$) (i.e. $\approx 10^{14} pseudo~ {\rm mK}^2h^{-3}{\rm Mpc^3} $) of the DATA and Model only roughly matches for the 14~m triads, however the same for 24~m and 28~m triads differ significantly. During the initial closure phase estimation stage, we found EoR0 data having multiple phase wraps, which could be due to the presence of some systematics or residual RFI. It caused an overall shift in the peak power away from zero delays in the coherent averaging. This effect can be seen in the 28~m triad, which shows greater power next to the zeroth delay (see figure \ref{fig: cross-power-EoR0}). The RMS floor level is between 1-2 orders of magnitude higher in the DATA ($\approx 10^{8}$ $pseudo~ {\rm mK}^2h^{-3}{\rm Mpc^3} $) compared to the Model ($\approx 10^{7}$ $ pseudo~ {\rm mK}^2h^{-3}{\rm Mpc^3} $) and Model with $\textbf{g}_{ij}$ ($\approx 10^{8}$ $ pseudo~ {\rm mK}^2h^{-3}{\rm Mpc^3} $). This excess power in the data compared to the Model may arise from a smaller DATA sample size in the EoR0 field or systematics and residual RFI. Using a baseline-dependent gain factor in the simulation, we aimed to incorporate such systematics. We performed a 2-sided KS test on the DATA and Model at $k_{||} > 0.15 ~pseudo ~h {\rm Mpc^{-1}}$, which shows rejection of the null hypothesis for the likelihood of DATA and Model drawn from the same distribution at all baseline cases, which is expected since both differ significantly. In contrast, the KS-test setifies the null hypothesis when comparing DATA and Model with $\textbf{g}_{ij}$.\par
    In the closure phase power spectrum of the EoR1 field, the peak power of the DATA and Model ($\approx 10^{15}$ $pseudo~ {\rm mK}^2h^{-3}{\rm Mpc^3}$) match for all triads. The RMS floor level between the Model and DATA gets significantly better compared to the EoR0 field. They nearly match in all cases, except for the 14~m triads where the difference is approximately an order of magnitude higher in the DATA compared to the Model (see figure \ref{fig: cross-power-EoR1}). It shows that we can improve our estimates of the power spectrum with an increased number of datasets. Thus, the analysis is data-limited for the amount of the data used. However, similar to the EoR0 case, the 2-sided KS test rejected the null hypothesis for all cases on Model, whereas fail to reject the null hypothesis in favour of the two distributions Model with $\textbf{g}_{ij}$ might be drawn from the same distribution.

    Since our observations lie in the middle of the DTV broadcasting band, we further investigated for residual RFI in our data. We shifted our entire analysis to the lower frequency band (167-177 MHz), avoiding the central DTV-affected band (although, as shown in figure no. 2 in \citep{Offringa_2015}, the DTV RFI nearly covers the entire EoR high band observations).
This analysis can help understand whether the residual RFI or other systematics are present in the data. Note that since we reduced our bandwidth by nearly three, we reduced our sensitivity by the same factor in the delay power spectrum. Therefore, the direct comparison of the mean power at $k_{||} > 0.15~pseudo~h {\rm Mpc^{-1}}$ might not be valid with the previous result.
 The closure phase power spectrum for the shifted spectrum is shown in figure \ref{fig: cross-power_shifted_EoR0}, and \ref{fig: cross-power_shifted_EoR1}. We can see significant improvements in the peak power of the DATA and Model compared to previous results. However, the overall RMS level was increased by an order of magnitude in all cases, possibly due to the lesser sensitivity (lower sample size). The DATA peak power at $\tau=0~\mu s$, especially in the EoR0 field, now matches the Model. Thus, we can justifiably argue that DTV RFI, which is expected to be prominent near 180 MHz, significantly contributed to the systematics present in the EoR0 DATA. On the other hand, EoR1 DATA seem relatively similar in both analyses, thus indicating that the systematics (such as persistent RFI) other than DTV RFI might be present in the data. Our findings are also confirmed when performing the KS-test on the DATA and Model, which shows non-rejection of the null hypothesis in all EoR1 field scenarios.
    We also compared the results with Model with $\textbf{g}_{ij}$ which are shown in table \ref{tab: KS-test-DTV} shows KS-test outcomes of the Data and Model and Data and Model with $\textbf{g}_{ij}$ on the shifted spectrum.
    
    We estimated the $2\sigma$ ($95\%$ confidence interval assuming Gaussianity from the convergence to Central Limit Theorem) on the 21-cm power spectrum for both the EoR0 and EoR1 fields. Our best upper limit on the 21-cm power spectrum of $\lesssim (184~pseudo~{\rm mK})^2$ came from EoR1 field on 14~m triads at $k=0.36~pseudo~h{\rm Mpc^{-1}}$ with the noise only uncertainties. In the EoR0 field, our best estimate, $\lesssim (188~pseudo~{\rm mK})^2$, came from the 24~m triads at $k=0.18~pseudo~h{\rm Mpc^{-1}}$ again using the noise-only uncertainty. However, as we discussed earlier, the systematics or residual RFI might have still affected these estimates, which we aim to address by introducing baseline-dependent gains in the modelling. It should be noted that, the exact nature of such baseline-dependent gains is not well understood. We have seen that, unlike Foregrounds, which usually gets restricted in lower delay modes, allowing faint HI signal to fluctuate visible at higher delay modes, the systematics equally affect all delay modes. Thus, for the scientific goal of observing milliradian-level sensitivity could be a significant challenge if such baseline dependent gains are present in the DATA. However, with extensive coherent averaging, the effect of such can be reduced. It should also be noted that the exact description of anomalies in the DATA can not be solely due to baseline-dependent systematics. Therefore, we state that if only the baseline-dependent systematics is present in the data, the level of noise introduced by the baseline-dependent gains justifies our DATA. Nevertheless, the level of fiducial H{\scriptsize{I}}~and FG+H{\scriptsize{I}}~powers are still lower than the data by $4-5$ and $3-4$ orders of magnitude, respectively; however, since our analysis is still data-limited, there is significant scope for improving upon the current estimates. Table~\ref{tab: 21-ps} provides the best $2\sigma$ estimates while table~\ref{tab: 21-ps_full} provides all estimates for each triad studied here.

\section{Summary}
We present independent EoR 21-cm power spectrum results from the closure phase analysis of $\approx 12$~hours of MWA phase-II data in the frequency range 167-197~MHz on three redundant baseline groups, namely, 14~m, 24~m, and 28~m baselines. Using the closure phase diagnostic, we found evidence for significant baseline-dependent systematics in the MWA data. Our best estimates of the 21-cm power spectrum at $z=6.79$ are $\lesssim (184~pseudo~{\rm mK})^2$ at $k=0.36$ $pseudo~h{\rm Mpc^{-1}}$ in the EoR1 field using 14~m baseline triads and $\lesssim (188~pseudo~{\rm mK})^2$ at $k=0.18$ $pseudo~h{\rm Mpc^{-1}}$ in the EoR0 field using 24~m baseline triads. Even with the limited amount of data analyzed, our closure phase method shows significant promise in independently constraining the 21-cm power spectrum during the EoR. Our results are still data-limited; hence, there is scope for further improvement by including more data in the analysis. Extensive sky modelling, such as accounting for the Galactic diffuse emission, is required before directly comparing the closure phase analysis with the standard visibility-based power spectrum. 

\section*{Acknowledgements} \label{section: acknowledgements}
This project is supported by an ARC Future Fellowship under grant FT180100321. This research was partially supported by the Australian Research Council Centre of Excellence for All Sky Astrophysics in 3 Dimensions (ASTRO 3D) through project number CE170100013. The International Centre for Radio Astronomy Research (ICRAR) is a Joint Venture of Curtin University and The University of Western Australia, funded by the Western Australian State government. The MWA Phase II upgrade project was supported by the Australian Research Council LIEF grant LE160100031 and the Dunlap Institute for Astronomy and Astrophysics at the University of Toronto. This scientific work uses Inyarrimanha Ilgari Bundara, the CSIRO Murchison Radio-astronomy Observatory. We acknowledge the Wajarri Yamatji people as the traditional owners and native title holders of the Observatory site. Support for the operation of the MWA is provided by the Australian Government (NCRIS) under a contract to Curtin University administered by Astronomy Australia Limited. We acknowledge the Pawsey Supercomputing Centre, which is supported by the Western Australian and Australian Governments. Data were processed at the Pawsey Supercomputing Centre. 

\section*{Data Availability} \label{section: data_availibility}
This project was developed using Python and dependent libraries mentioned in the main text. Our data processing pipeline is publically available at \texttt{Github}\footnote{\url{https://github.com/himmng/Closure_phase_analysis}}, along with the final processed datasets. The core data used in this work will be made available upon reasonable request.


\printbibliography

\appendix

\section{Appendix} \label{section: appendix}
\subsection{Inner vs Outer triads} \label{inner_vs_outer}
\begin{figure*}
    \centering
    \includegraphics[scale=0.45]{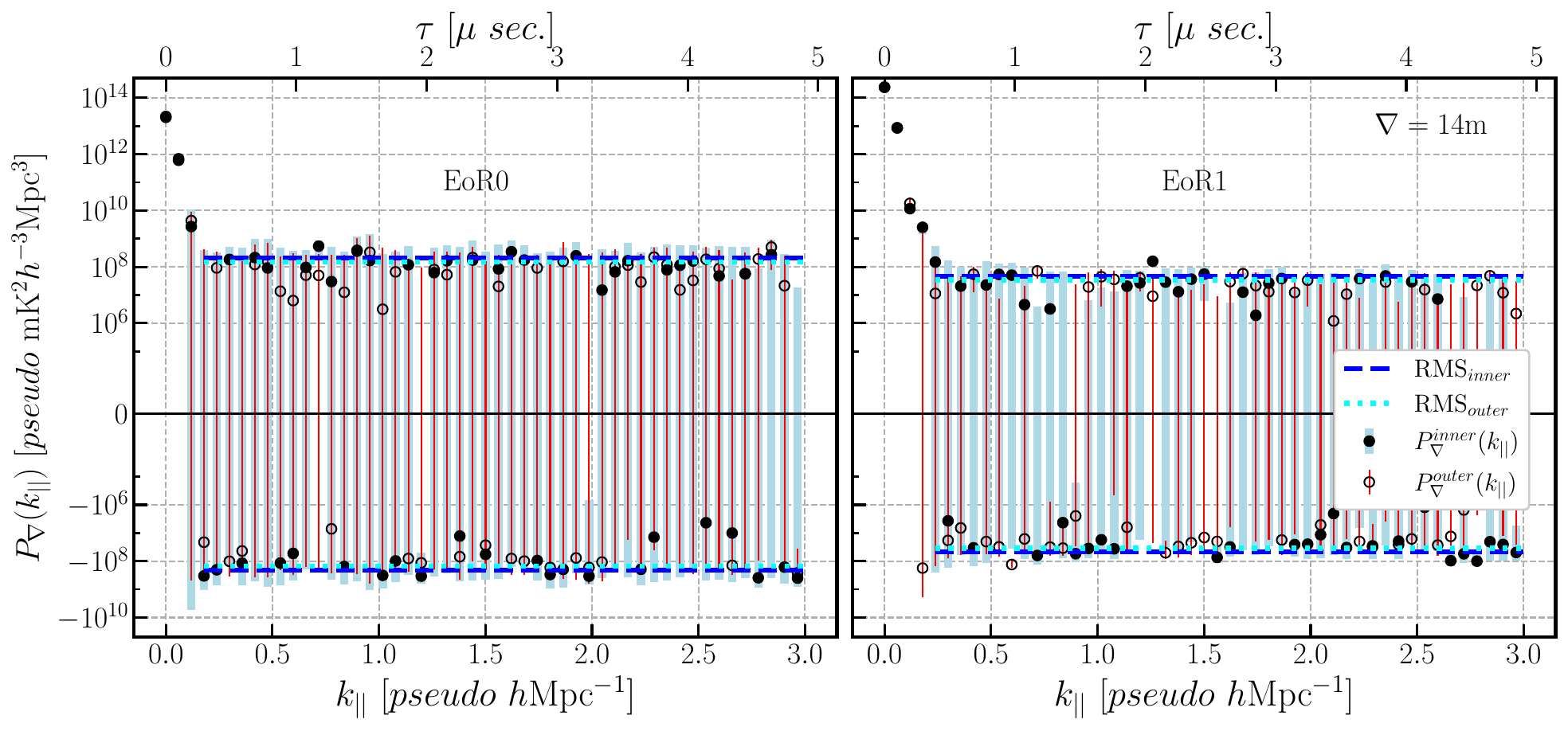}
    \caption{Power spectrum comparison between the Inner (filled circles) and Outer (empty circles) in 14~m triads for EoR0 (\textit{Left}) and EoR1 (\textit{Right}) fields, respectively.}
    \label{fig: pow_inner_vs_outer}
\end{figure*}

The core of MWA Hexagons may be more affected by mutual coupling than the tiles near the edges. Therefore, we tried to perform an independent check on the power differences between the inner and outer tile triads to quantify the cross-talk level in the data. We chose $14$m baseline triads to get a higher count and only estimated the closure phase power spectrum from them. We used first and second outer layer of the Hexagon configuration as the outer triads, and third and fourth tile layers as inner triads (see right panel of figure \ref{fig: MWAII_config}). In scenarios, where either two tiles from outer while the third from inner, or vice-versa, we considered those triads to be outer or inner triads, respectively. Figure \ref{fig: pow_inner_vs_outer} shows the closure phase power spectrum of the inner and outer triads at $14$m baseline lengths. We observed the relative percentage difference between the RMS power estimated for $k_{||}>0.15 ~pseudo ~h\,\rm Mpc^{-1}$ between inner triads is higher than the outer triads are about $41\%, 34\%$ for EoR0 and EoR1 fields, respectively. However, the relative difference with the Model RMS (figure \ref{fig: cross-power_shifted_EoR0}, \ref{fig: cross-power_shifted_EoR1}) $\approx$ 6\%, -25\% for inner and outer triads in EoR0 field, and $\approx$ 186\%, 114\% for inner and outer triads in EoR1 field. These when comparing with the mean RMS value of the Model ($\approx 10^8, 10^7 pseudo~ {\rm mK}^2h^{-3}{\rm Mpc^3}$ for E0R0, EoR1 fields) consistent with each other.

\subsection{Avoiding DTV}
The observations are centred around 180 MHz, which overlap with the Australian Digital Television (DTV) band; therefore, despite the extensive RFI flagging, some RFI could remain in the final processed data. Therefore, a useful test would be to shift the spectral window to lower frequencies, avoiding the DTV band in the analysis.  

\begin{figure*}
    \centering
    \includegraphics[scale=0.35]{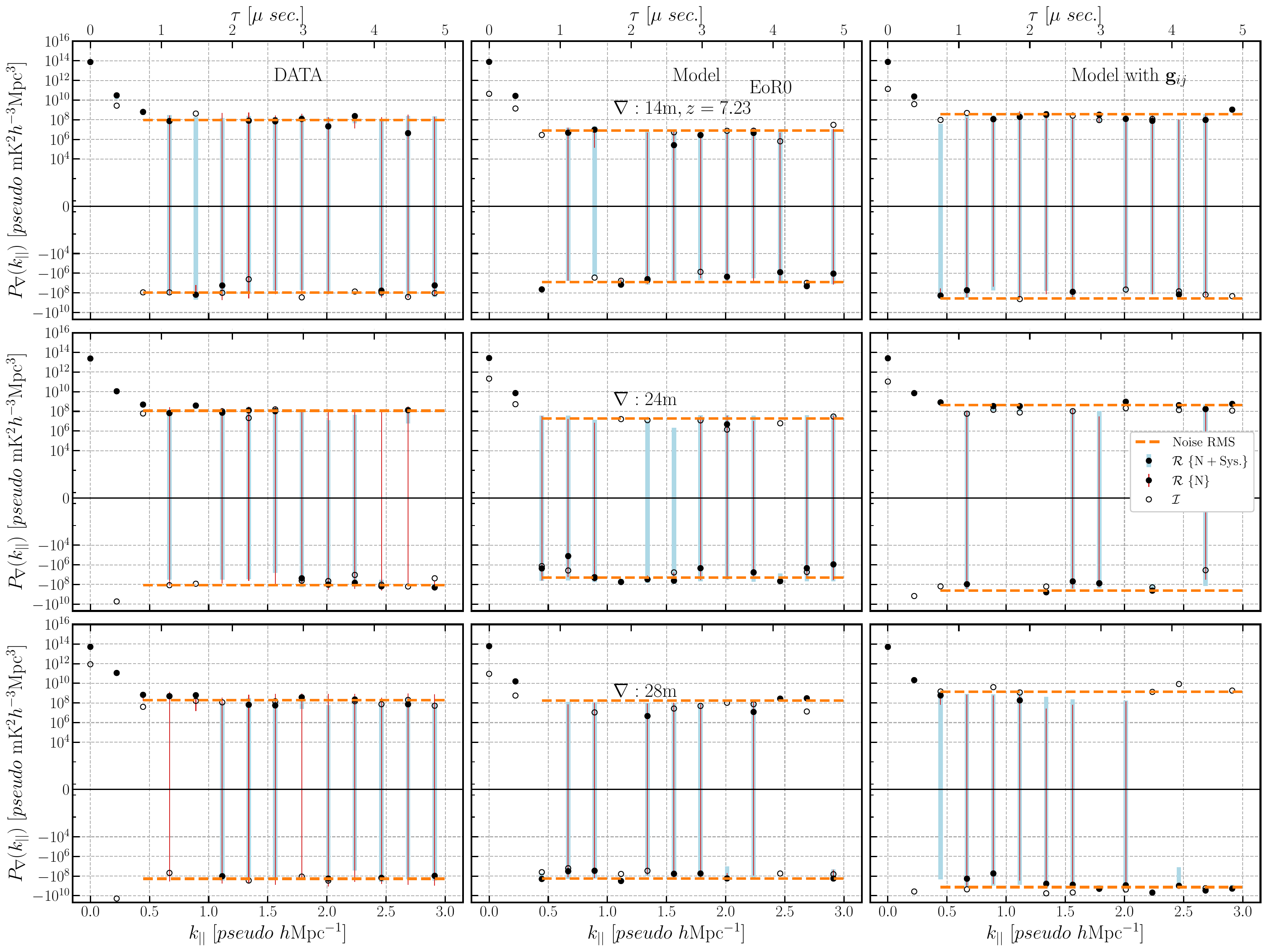}
    \caption{Cross power spectrum of the closure phase delay spectrum for EoR0 observing field when the window function is shifted towards lower frequencies (167-177) MHz to avoid the DTV frequency band around 180 MHz. All symbols, colors, and line styles are the same as in figure~\ref{fig: cross-power-EoR0}.}
    \label{fig: cross-power_shifted_EoR0}
\end{figure*}

\begin{figure*}
    \centering
    \includegraphics[scale=0.35]{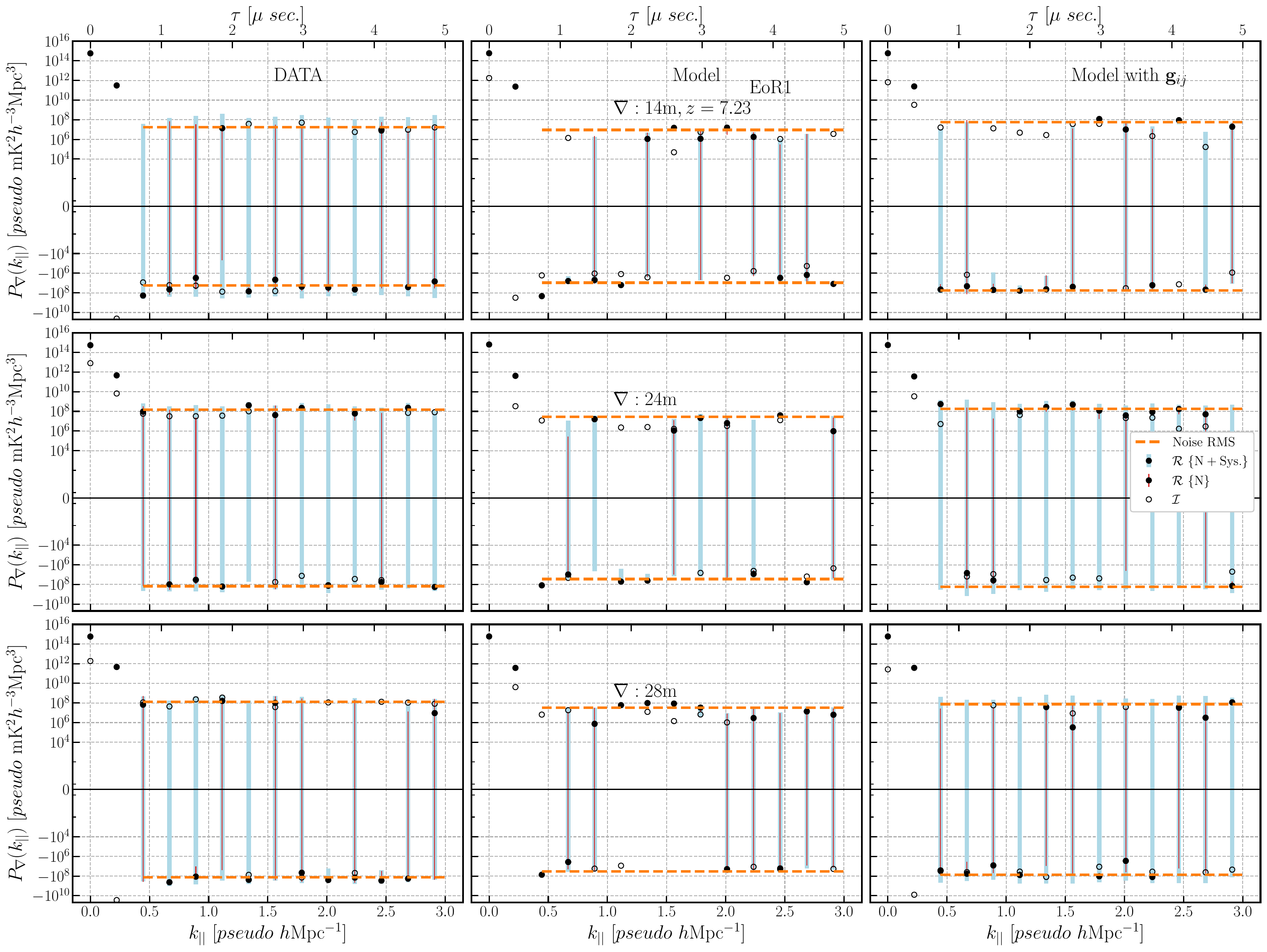}
    \caption{Same as fig.\ref{fig: cross-power_shifted_EoR0} but for the EoR1 field.}
    \label{fig: cross-power_shifted_EoR1}
\end{figure*}

To do this, we worked with the first $\approx 10$ MHz band (167--177~MHz) of the total 30.72~MHz bandwidth and estimated the cross-power spectrum using the same procedure. However, the effective bandwidth for this analysis was reduced to $\approx 3.3$ MHz, thus reducing the sensitivity by the same factor. 
The idea was to check whether the floor level of the power spectra (refer to figure \ref{fig: cross-power-EoR0}, \ref{fig: cross-power-EoR1}), which shows the excess power in the DATA, reduces and matches with the Model in the shifted window. It would suggest that the RFI could be localised in the central portion of the band, which is one of the major contributors to the excess power.
However, the power spectra of shifted window (see figure~\ref{fig: cross-power_shifted_EoR0} and \ref{fig: cross-power_shifted_EoR1}) show similar behavior as that in figure~\ref{fig: cross-power-EoR0} and \ref{fig: cross-power-EoR1}, indicating that the level of the RFI might be ubiquitous across the entire observing band along with the systematics in the data which are not being modelled in the simulations.

\begin{table}
\centering
\caption{2-sided KS test outcomes on DTV avoided band. The null hypothesis compares the Data and Model, Data and Model with $\mathbf{g}_{ij}$ at $k_{||}>0.15~ [pseudo ~h\,\rm Mpc^{-1}]$.}
\begin{tabular}{|c|c|cc|ll|} 
    \hline
  \multicolumn{6}{|c|}{2-sided KS-test}\\ \hline
 field&$\nabla$& \multicolumn{2}{c|}{Model}& \multicolumn{2}{c|}{Model with $\mathbf{g}_{ij}$}\\ \hline
  && $p$-value & statistic& $p$-value&statistic\\ 

     & $14$~m & \fbox{$0.04$}&$0.54$&  \colorbox{gray}{$0.98$}&$0.15$\\ 
EoR0 & $24$~m & \fbox{$0.01$}&$0.61$& \colorbox{gray}{$0.30$}&$0.38$\\ 
     & $28$~m &  \fbox{$0.04$}& $0.54$&  \colorbox{gray}{$0.13$}&$0.46$\\ \hline
          & $14$~m & \colorbox{gray}{$0.13$}& $0.46$& \colorbox{gray}{$0.90$}&$0.23$\\ 
EoR1 & $24$~m & \colorbox{gray}{$0.59$}& $0.30$& \colorbox{gray}{$0.30$}&$0.38$\\ 
     & $28$~m &  \colorbox{gray}{$0.13$}& $0.46$& \colorbox{gray}{$0.13$}&$0.46$\\ \hline
\end{tabular}\label{tab: KS-test-DTV}

\end{table}


\subsection{Diagnosing bandpass systematics}
We checked the closure phases for the baseline-dependent systematics, which persisted in the MWA data by modifying the antenna element-based gains ($\mathbf{g}_i'\rm s$ eq. \ref{eq: closure_phase}).  
First, we created one set of MWA bandpass using random Gaussian between [0, 1] for each edge-channel frequency. We multiplied these by each visibilities in the correct parity pair order. It provides us with modified visibilities with randomised gains at the edge channels, which mimics the bandpass-affected visibilities. We used simulated flux densities for this procedure since they were produced ideally with unity antenna gains (although it does not affect having unity gains in the first place). 

\begin{figure*}
    \centering
    \includegraphics[scale=0.45]{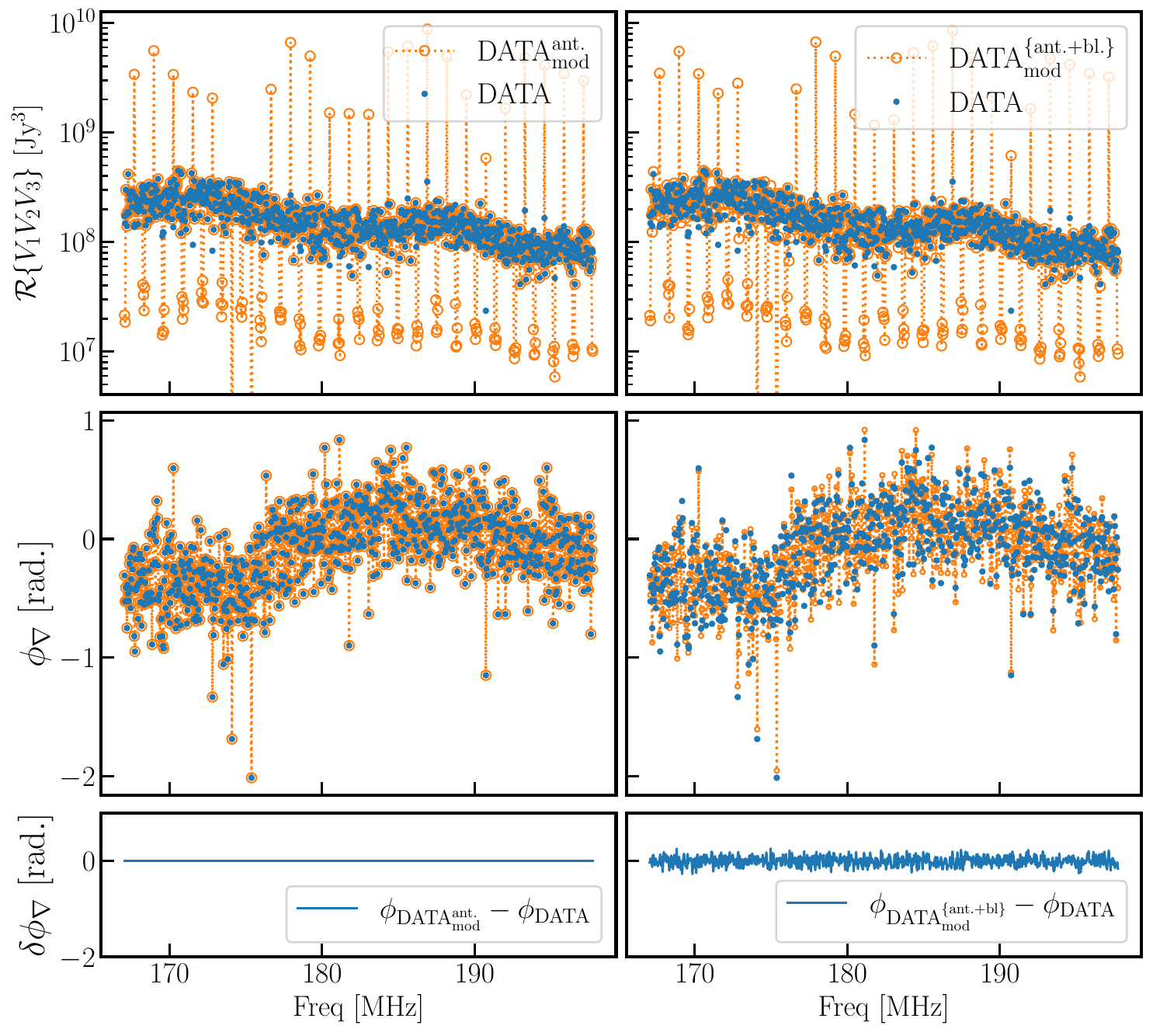}
    \caption{Antenna-element gains and baseline-dependent gains inserted in the data at 1.28 MHz regular intervals. The top-bottom panel shows the bispectrum, closure phase, and residual closure phase between the data and modified data. \textit{Left}: Showing antenna-element gains getting eliminated in the closure phase. \textit{Right}: baseline-dependent gains persist in the closure phase.}
    \label{fig: bandpass-sim}
\end{figure*}

Second, we used two scenarios to modify the bandpass further. In the first, the antenna-element-based gains were modified with new gains multiplied by the existing ones. This step verifies how the individual antenna-based gains vanish in the closure phase, illustrated in the top panel of figure \ref{fig: bandpass-sim}. In the second scenario, instead of multiplicative gains from individual antenna elements (e.g. $\mathbf{g}_i, \mathbf{g}_j$), we multiplied an additional baseline-dependent term ($\mathbf{g}_{ij}$) that is not factorisable into element-based terms. It demonstrated that baseline-dependent gains do not cancel in the closure phase; see the same figure~\ref{fig: bandpass-sim}  bottom panel, where the residual difference between the residuals between the closure phase modified with $\mathbf{g}_{ij}$ and the original does not vanish.

\subsection{Incomplete modelling}
\begin{figure*}
    \centering
    \includegraphics[scale=0.44]{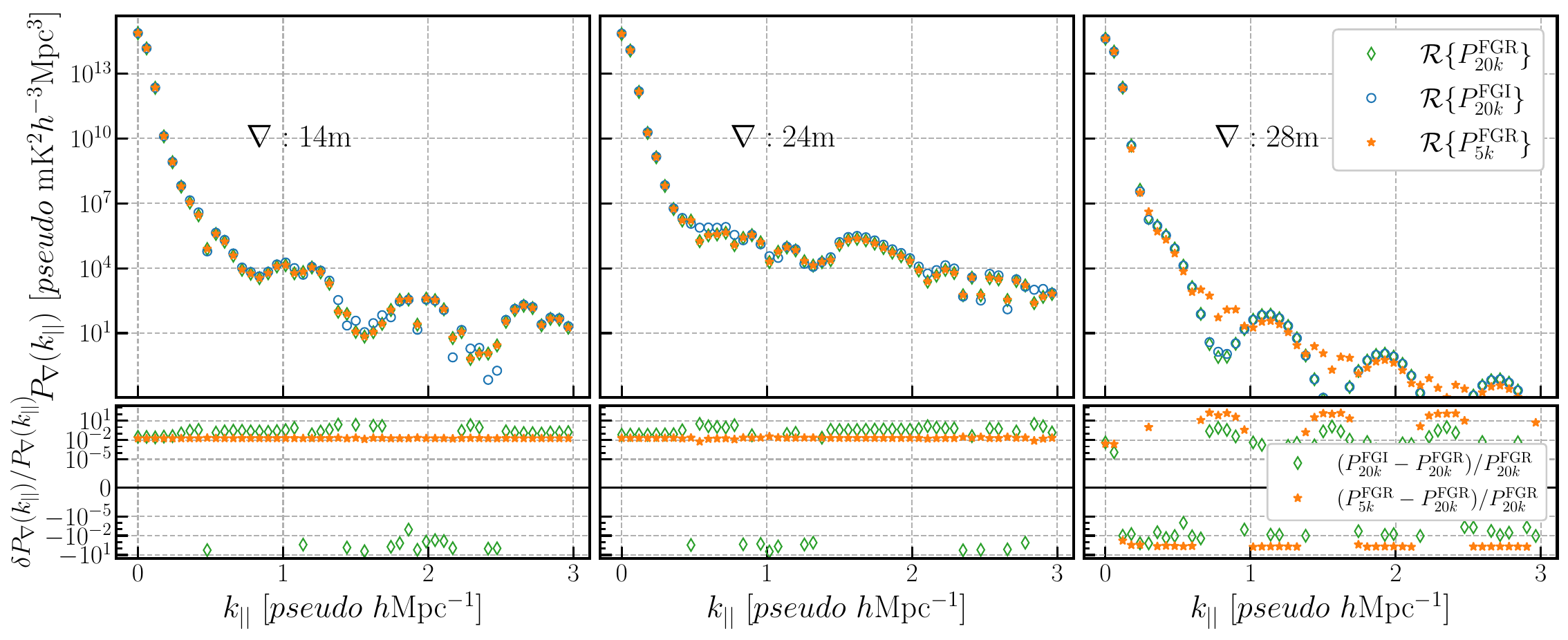}
    \caption{Top: Foreground model power for three scenarios, Model with 20,000 sources with real dipole gains, Model with 20,000 sources with unity (\textit{ideal}) dipole gains, foreground model with 5,000 sources with real dipole gains in EoR1 field. Bottom: The relative difference between the unity dipole gains model (20,000 sources), and real dipole gains (5,000 sources) at 14~m, 24~m, 28~m with real dipole gains model (20,000 sources).}
    \label{fig: FG-compare2}
\end{figure*}

Realistic vs. ideal beams or using fewer foreground sources can affect the final power estimates. Thus, we produced two test foreground simulations, one with the same 20,000 foreground sources but with ideal beam conditions where all dipole gains are active and set to unity ($\rm FGI_{20k}$) and a second with only 5,000 foreground sources and real beam conditions, dead/missing dipoles incorporated in the beam evaluation ($\rm FGR_{5k}$), which we compared against 20,000 foreground sources but with real beam conditions ($\rm FGR_{20k}$).
Note that in the main results, we used 20,000 foreground sources with real beams, which are compared against the two test scenarios. The three cases' final closure phase power spectrum, foreground with real dipole gains with 20,000 sources,  foreground with unity dipole gains with 20,000 sources, and foreground with real dipole gains with 5,000 sources, are shown in figure \ref{fig: FG-compare2}. We obtained RMS power at $\geq 1.0 ~pseudo~h\rm Mpc^{-1}$ to differentiate the two scenarios with the foreground with real dipole gains with 20,000 sources. The relative percentage error for unity dipole gains (20,000 sources) at $14$m, $24$m, and $28$m baselines are $4\%, 39\%, 0.3\%$, and for real dipole gains (5,000 sources) at $14$m, $24$m, $28$m baselines are $1.7\%, 1.9\%, 40\%$, respectively. Comparing with the RMS value of the Model from figure \ref{fig: cross-power_shifted_EoR0}, \ref{fig: cross-power_shifted_EoR1}) ($\approx 10^8, 10^7 pseudo~ {\rm mK}^2h^{-3}{\rm Mpc^3}$ for E0R0, EoR1 fields), these relative differences in both the scenarios are sufficiently less, thus, consistent with the Model.

\subsection{Data Structure Flowchart}

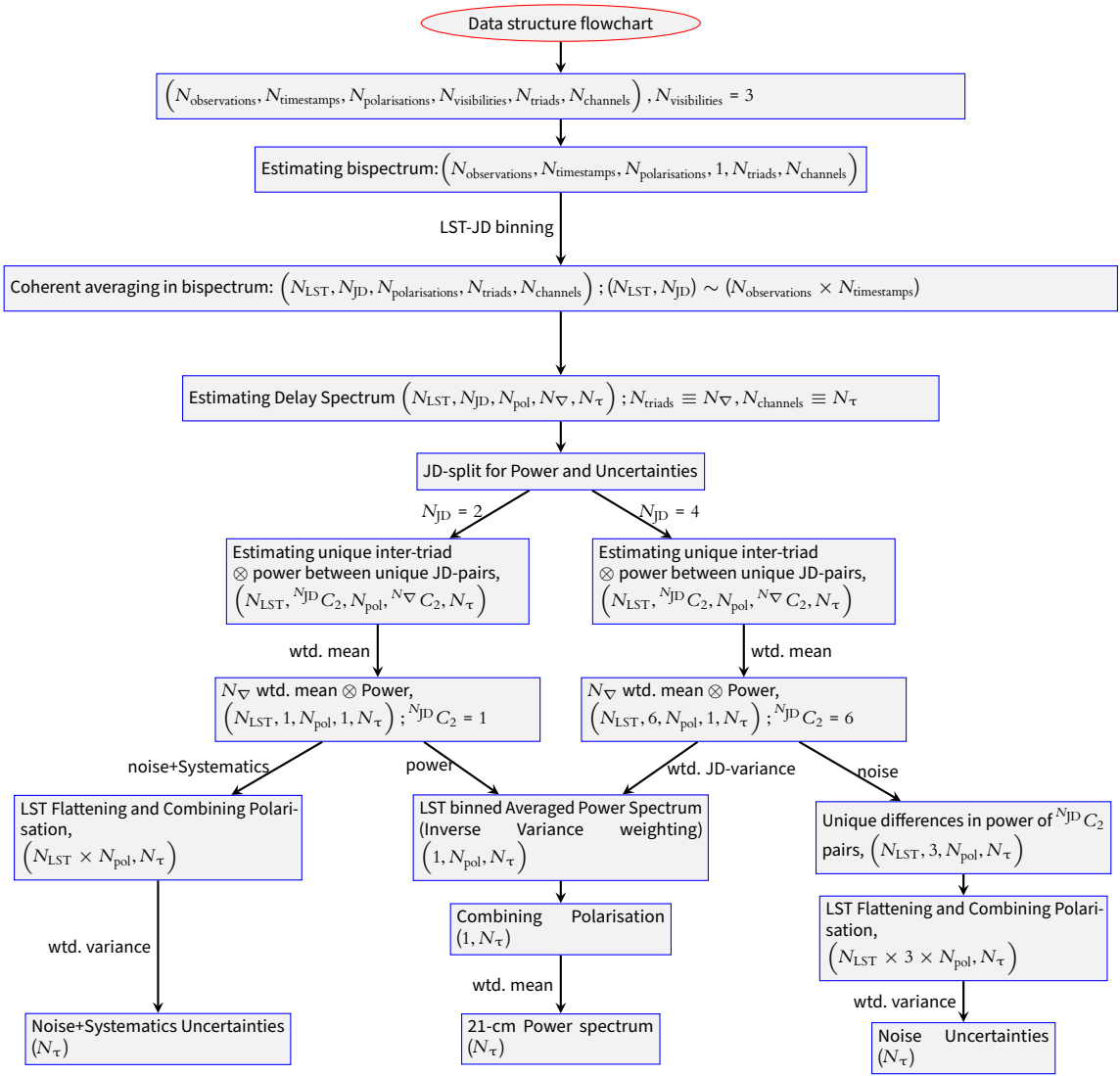
\begin{figure*}
\caption{Schematic flow chart of the data structure through processing pipeline}
\centering

{\scriptsize \begin{tikzpicture}[node distance=2em]

\node (S) [startend] {Data structure flowchart};

\node (S1) [process, below of=S, yshift= -2em]{\parbox[t][][t]{10.7cm}{
 $\left( N_{\rm observations}, N_{\rm timestamps},  N_{\rm polarisations}, N_{\rm visibilities}, N_{\rm triads}, N_{\rm channels}\right), N_{\rm visibilities}=3$}};

\node (S2) [process, below of=S1, yshift=-2em]{Estimating bispectrum:$\left( N_{\rm observations}, N_{\rm timestamps},  N_{\rm polarisations}, 1, N_{\rm triads}, N_{\rm channels}\right)$};

\node (S3) [process, below of=S2, yshift=-4.5em]{\parbox[t][][t]{14.8cm}{Coherent averaging in bispectrum: $\left( N_{\rm LST}, N_{\rm JD},  N_{\rm polarisations}, N_{\rm triads}, N_{\rm channels}\right); (N_{\rm LST}, N_{\rm JD}) \sim (N_{\rm observations}\times N_{\rm timestamps})$}};

\node (S4) [process, below of=S3, yshift=-4em]{\parbox[t][][t]{10.cm}{Estimating Delay Spectrum $\left( N_{\rm LST}, N_{\rm JD},  N_{\rm pol}, N_{\rm \nabla}, N_{\tau}\right); N_{\rm triads}\equiv N_\nabla, N_{\rm channels}\equiv N_\tau$}};

\node (S5) [process, below of=S4, yshift=-2em]{\parbox[t][][t]{3.7cm}{JD-split for Power and Uncertainties}};

\node (S6) [process, below of=S5, yshift=-4em, xshift=-10em]{\parbox[t][][t]{3.9cm}{ Estimating unique inter-triad \\
$\otimes$ power between unique JD-pairs,\\
$\left( N_{\rm LST}, \prescript{N_{\rm JD}}{}{C}_{2},  N_{\rm pol}, \prescript{N_{\nabla}}{}{C}_{2}, N_{\tau}\right)$}};

\node (S6_) [process, below of=S5, yshift=-4em, xshift=10em]{\parbox[t][][t]{3.9cm}{ Estimating unique inter-triad \\
$\otimes$ power between unique JD-pairs,\\
$\left( N_{\rm LST}, \prescript{N_{\rm JD}}{}{C}_{2},  N_{\rm pol}, \prescript{N_{\nabla}}{}{C}_{2}, N_{\tau}\right)$}};

\node (S7) [process, below of=S6, yshift=-5em,]{\parbox[t][][t]{4.2cm}{$N_\nabla$ wtd. mean $\otimes$ Power,\\ $\left( N_{\rm LST}, 1,  N_{\rm pol}, 1, N_{\tau}\right); \prescript{N_{\rm JD}}{}{C}_{2}=1$}};

\node (S7_) [process, below of=S6_, yshift=-5em]{\parbox[t][][t]{4.2cm}{$N_\nabla$ wtd. mean $\otimes$ Power,\\ $\left( N_{\rm LST}, 6,  N_{\rm pol}, 1, N_{\tau}\right); \prescript{N_{\rm JD}}{}{C}_{2}=6$}};

\node (S8) [process, below of=S7, yshift=-5em, xshift=10em]{\parbox[t][][t]{3.8cm}{ LST binned Averaged Power Spectrum\\ (Inverse Variance weighting) $\left( 1,N_{\rm pol}, N_{\tau}\right)$}};

\node (S9) [process, below of=S8, yshift=-3em]{\parbox[t][][t]{2.8cm}{Combining Polarisation $\left( 1, N_{\tau}\right)$}};

\node (S10) [process, below of=S9, yshift=-4em]{\parbox[t][][t]{2.5cm}{21-cm Power spectrum $\left(N_{\tau}\right)$}};

\node (S11) [process, below of=S7_, yshift=-5em, xshift=12em]{\parbox[t][][t]{3.8cm}{Unique differences in power of $\prescript{N_{\rm JD}}{}{C}_{2}$ pairs, $\left( N_{\rm LST}, 3,  N_{\rm pol}, N_{\tau}\right)$}};

\node (S12) [process, below of=S11, yshift=-3.5em]{\parbox[t][][t]{3.7cm}{LST Flattening and Combining Polarisation,\\ $\left( N_{\rm LST}\times 3 \times N_{\rm pol}, N_{\tau}\right)$}};

\node (S13) [process, below of=S12, yshift=-4em]{\parbox[t][][t]{2.3cm}{Noise Uncertainties $\left(N_{\tau}\right)$}};

\node (S14) [process, below of=S7, yshift=-5em, xshift=-12em]{\parbox[t][][t]{3.7cm}{LST Flattening and Combining Polarisation,\\ $\left( N_{\rm LST}\times  N_{\rm pol}, N_{\tau}\right)$}};

\node (S15) [process, below of=S14, yshift=-9em]{\parbox[t][][t]{3.4cm}{Noise+Systematics Uncertainties $\left(N_{\tau}\right)$}};

\draw [arrow] (S) -- (S1);
\draw [arrow] (S1) -- (S2);
\draw [arrow] (S2) -- node[anchor=east]{LST-JD binning}(S3);
\draw [arrow] (S3) -- (S4);
\draw [arrow] (S4) -- (S5);
\draw [arrow] (S5) -- node[anchor=east]{$N_{\rm JD}=2$}(S6);
\draw [arrow] (S5) -- node[anchor=west]{$N_{\rm JD}=4$}(S6_);
\draw [arrow] (S6) -- node[anchor=east]{wtd. mean }(S7);
\draw [arrow] (S6_) -- node[anchor=west]{wtd. mean}(S7_);
\draw [arrow] (S7_) -- node[anchor=west]{wtd. JD-variance}(S8);
\draw [arrow] (S7) -- node[anchor=east]{power}(S8);
\draw [arrow] (S8) -- (S9);
\draw [arrow] (S9) -- node[anchor=east]{wtd. mean}(S10);

\draw [arrow] (S7_) -- node[anchor=west]{noise}(S11);
\draw [arrow] (S11) -- (S12);
\draw [arrow] (S12) -- node[anchor=east]{wtd. variance}(S13);

\draw [arrow] (S7) -- node[anchor=east]{noise+Systematics}(S14);
\draw [arrow] (S14) -- node[anchor=east]{wtd. variance}(S15);

\end{tikzpicture}} 

\label{fig: data_flowchart}
\end{figure*}

As we proceed through the different averaging steps, the data structure is shown as a flowchart in figure~\ref{fig: data_flowchart}.

\begin{table*}
\caption{Complete table of 2$\sigma$ upper limit estimates of 21-cm power spectrum [\textit{pseudo} ${\rm mK^2}$] }
\resizebox{18cm}{!}{
\begin{threeparttable}
\begin{tabular}{||l||c|c|c|c|c|c|c|c|c|c|c|c|} \hline 
  \multicolumn{13}{|c|}{$\Delta_{\nabla \rm ~UL}^2 [pseudo~ {\rm mK^2}]$}\\ 
\hline
Field
   &\multicolumn{6}{c|}{EoR0}&  \multicolumn{6}{c|}{EoR1}\\ \hline
Baseline
&\multicolumn{2}{c|}{$\nabla: 14~\rm m$} 
&\multicolumn{2}{c|}{$\nabla: 24~\rm m$}
&\multicolumn{2}{c|}{$\nabla: 28~\rm m$}
&\multicolumn{2}{c|}{$\nabla: 14~\rm m$}
&\multicolumn{2}{c|}{$\nabla: 24~\rm m$}
&\multicolumn{2}{c|}{$\nabla: 28~\rm m$}\\\hline
$k$ [\textit{pseudo}~$h\,{\rm Mpc^{-1}}$]
& N. & N.+Sys.
& N. & N.+Sys.
& N. & N.+Sys.
& N. & N.+Sys.
& N. & N.+Sys.
& N. & N.+Sys.\\
\hline\hline

$0.18$&-- & $(392^*)^2$&$(188^*)^2$ & $(207^*)^2$&-- & --&-- & $(526)^2$&-- & $(361)^2$&-- & --\\
$0.24$&$(347^*)^2$ & $(420^*)^2$&-- & --&-- & --&-- & $(427)^2$&-- & $(458)^2$&$(236)^2$ & $(314)^2$\\
$0.30$&-- & $(534^*)^2$&-- & --&-- & --&$(218)^2$ & $(263)^2$&-- & $(503)^2$&-- & $(512)^2$\\
$0.36$&$(490^*)^2$ & $(608^*)^2$&-- & --&-- & --&$(184)^2$ & $(330)^2$&-- & $(849)^2$&-- & $(572)^2$\\
$0.42$&-- & $(1562^*)^2$&-- & $(1065^*)^2$&$(732^*)^2$ & $(708^*)^2$&$(474)^2$ & $(434)^2$&-- & $(1037)^2$&-- & $(762)^2$\\
$0.48$&-- & $(2391)^2$&$(1133)^2$ & $(1358)^2$&$(834)^2$ & $(943)^2$&-- & --&-- & $(1888)^2$&-- & $(1082)^2$\\
$0.54$&-- & $(1593)^2$&$(658)^2$ & $(772)^2$&-- & --&-- & --&$(1252)^2$ & $(1336)^2$&-- & $(998)^2$\\
$0.60$&$(1152)^2$ & $(1525)^2$&-- & --&-- & --&-- & --&$(1531)^2$ & $(2013)^2$&-- & $(1129)^2$\\
$0.66$&$(1329)^2$ & $(1830)^2$&$(724)^2$ & $(932)^2$&$(826)^2$ & $(989)^2$&$(687)^2$ & $(720)^2$&-- & $(1610)^2$&-- & $(1639)^2$\\
$0.72$&-- & $(2298)^2$&-- & $(1098)^2$&$(1436)^2$ & $(1539)^2$&-- & --&$(1083)^2$ & $(2148)^2$&-- & $(1875)^2$\\
$0.78$&$(1252)^2$ & $(2312)^2$&-- & --&-- & --&$(785)^2$ & $(855)^2$&$(958)^2$ & $(2090)^2$&$(1203)^2$ & $(1509)^2$\\
$0.84$&$(1407)^2$ & $(2284)^2$&$(1660)^2$ & $(1771)^2$&-- & --&-- & $(884)^2$&-- & $(2008)^2$&-- & $(2902)^2$\\
$0.90$&-- & $(7324)^2$&-- & --&-- & --&-- & $(1391)^2$&-- & $(3798)^2$&$(1193)^2$ & $(2520)^2$\\
$0.96$&-- & $(8728)^2$&$(3935)^2$ & $(4599)^2$&-- & --&$(1325)^2$ & $(1647)^2$&-- & $(6085)^2$&$(598)^2$ & $(2326)^2$\\
$1.02$&$(3608)^2$ & $(5318)^2$&$(2004)^2$ & $(2214)^2$&-- & --&$(1045)^2$ & $(1325)^2$&-- & $(3104)^2$&-- & $(1753)^2$\\
$1.08$&-- & $(5048)^2$&-- & --&$(1961)^2$ & $(2473)^2$&$(1024)^2$ & $(1212)^2$&-- & $(4041)^2$&-- & $(2176)^2$\\
$1.14$&$(3273)^2$ & $(4149)^2$&-- & $(3152)^2$&-- & --&$(1375)^2$ & $(1648)^2$&-- & $(4893)^2$&$(1327)^2$ & $(2708)^2$\\
$1.20$&-- & $(3694)^2$&-- & --&$(2501)^2$ & $(2524)^2$&$(1828)^2$ & $(1758)^2$&$(2342)^2$ & $(3550)^2$&-- & $(2273)^2$\\
$1.26$&$(2841)^2$ & $(5036)^2$&-- & --&-- & --&-- & --&$(1038)^2$ & $(4967)^2$&$(2468)^2$ & $(3158)^2$\\
$1.32$&$(5160)^2$ & $(6225)^2$&$(2393)^2$ & $(2832)^2$&$(2412)^2$ & $(2996)^2$&-- & --&$(1360)^2$ & $(5115)^2$&$(1885)^2$ & $(3286)^2$\\
$1.38$&$(6337)^2$ & $(8341)^2$&-- & $(7501)^2$&-- & --&$(1393)^2$ & $(1788)^2$&-- & $(9528)^2$&-- & $(3420)^2$\\
$1.44$&-- & --&$(4462)^2$ & $(4316)^2$&$(5041)^2$ & $(5977)^2$&-- & $(2156)^2$&$(3262)^2$ & $(6583)^2$&-- & $(3391)^2$\\
$1.50$&$(5275)^2$ & $(7175)^2$&$(2466)^2$ & $(3263)^2$&$(2525)^2$ & $(3751)^2$&$(2926)^2$ & $(2748)^2$&$(739)^2$ & $(3695)^2$&$(3758)^2$ & $(4835)^2$\\
$1.56$&$(5722)^2$ & $(6703)^2$&$(3258)^2$ & $(3487)^2$&-- & --&-- & $(3149)^2$&$(3273)^2$ & $(4824)^2$&-- & $(5720)^2$\\
$1.62$&$(6298)^2$ & $(6931)^2$&-- & --&-- & --&$(2313)^2$ & $(2826)^2$&-- & $(5113)^2$&$(2798)^2$ & $(4530)^2$\\
$1.68$&$(5693)^2$ & $(7257)^2$&-- & --&$(3419)^2$ & $(3759)^2$&$(2551)^2$ & $(2965)^2$&-- & $(7644)^2$&-- & $(4205)^2$\\
$1.74$&-- & $(6169)^2$&-- & --&-- & $(4612)^2$&$(1920)^2$ & $(2523)^2$&-- & $(3926)^2$&-- & $(5890)^2$\\
$1.80$&-- & $(10763)^2$&$(5303)^2$ & $(5298)^2$&-- & $(10595)^2$&$(2145)^2$ & $(2714)^2$&$(3837)^2$ & $(7290)^2$&$(4584)^2$ & $(5548)^2$\\
$1.86$&-- & $(19820)^2$&-- & --&-- & --&$(3223)^2$ & $(3471)^2$&-- & $(15979)^2$&-- & $(9441)^2$\\
$1.92$&-- & $(12907)^2$&$(6168)^2$ & $(7035)^2$&-- & --&$(3953)^2$ & $(3897)^2$&$(3023)^2$ & $(7422)^2$&-- & $(9452)^2$\\
$1.99$&-- & --&$(4682)^2$ & $(5501)^2$&$(7153)^2$ & $(7474)^2$&-- & $(4728)^2$&-- & $(8259)^2$&$(3951)^2$ & $(6392)^2$\\
$2.05$&-- & $(12125)^2$&-- & --&-- & --&$(3891)^2$ & $(4085)^2$&-- & $(8977)^2$&$(2479)^2$ & $(7569)^2$\\
$2.11$&$(6947)^2$ & $(11203)^2$&-- & --&-- & $(7096)^2$&-- & $(4641)^2$&-- & $(5832)^2$&-- & $(8127)^2$\\
$2.17$&$(8365)^2$ & $(8825)^2$&$(4205)^2$ & $(5379)^2$&-- & $(7607)^2$&-- & --&$(6522)^2$ & $(9015)^2$&-- & --\\
$2.23$&-- & $(13263)^2$&-- & --&$(6100)^2$ & $(6331)^2$&-- & --&$(5587)^2$ & $(10802)^2$&-- & --\\
$2.29$&-- & $(16087)^2$&-- & --&$(6661)^2$ & $(8300)^2$&$(5563)^2$ & $(4660)^2$&$(4315)^2$ & $(8076)^2$&-- & $(9638)^2$\\
$2.35$&-- & $(17405)^2$&$(7399)^2$ & $(8814)^2$&$(5434)^2$ & $(6674)^2$&$(5612)^2$ & $(6196)^2$&-- & $(14833)^2$&-- & $(11609)^2$\\
$2.41$&$(9506)^2$ & $(13119)^2$&$(7532)^2$ & $(8437)^2$&$(8020)^2$ & $(9034)^2$&$(4847)^2$ & $(5121)^2$&-- & $(11582)^2$&-- & $(11719)^2$\\
$2.47$&$(11678)^2$ & $(14416)^2$&-- & --&$(5505)^2$ & $(7577)^2$&$(4176)^2$ & $(4897)^2$&$(6363)^2$ & $(10704)^2$&$(3984)^2$ & $(8565)^2$\\
\hline\hline
\end{tabular}
\begin{tablenotes}
   \item[a]  k-modes where the uncertainty brackets do not include zero power are masked and shown with dashes ($-$).
   \item[b] limits quoted with an asterisk ($*$) might be affected by systematics or persistent residual RFI.
\end{tablenotes}
\end{threeparttable}}
\end{table*} \label{tab: 21-ps_full}

\end{document}